\documentclass[10pt]{LMA}
\usepackage{comment}

\usepackage{multirow}
\title{Multi-step phase transitions and gravitational waves in the inert doublet model}
\author[a]{Nico Benincasa\thanks{nico.benincasa@kbfi.ee; corresponding author}}
\author[b]{Luigi Delle Rose\thanks{luigi.dellerose@unical.it}}
\author[a]{Kristjan Kannike\thanks{kristjan.kannike@cern.ch}}
\author[a]{Luca Marzola\thanks{luca.marzola@cern.ch}}
\affil[a]{National Institute of Chemical Physics and Biophysics,  R\"avala 10, Tallinn, Estonia}
\affil[b]{Dipartimento di Fisica, Università della Calabria, I-87036 Arcavacata di Rende, Cosenza, Italy}

\date{\today}

\begin{document}

\maketitle

\begin{abstract}
The inert doublet model is a well-motivated extension of the Standard Model that contains a dark matter candidate and  modifies the dynamics of the electroweak symmetry breaking. In order to detail its phenomenology, we perform a comprehensive study of cosmic phase transitions and gravitational wave signals implied by the framework, accounting for the latest results of collider experiments. We require the neutral inert scalar to constitute, at least, a subdominant part of the observed dark matter abundance. While most of the phase transitions proceed through a single step, we identify regions of the parameter space where the electroweak vacuum is reached after multiple phase transitions. The resulting gravitational wave spectrum is generally dominated by single-step transitions and, in part of the parameter space, falls within the reach of future gravitational wave detectors such as LISA, BBO or DECIGO. We find that direct detection experiments efficiently probe the part of parameter space associated with multi-step phase transitions, which remain unconstrained only in the Higgs resonance region testable with future monojet searches. The implications of the new determination of the $W$ boson mass are also discussed. 
\end{abstract}

\section{Introduction}
\label{sec:intro}

Although the discovery of the Higgs boson at the LHC~\cite{CMS:2012qbp,ATLAS:2012yve}  brought to completion the search for Standard Model (SM) particles, 
we are far from having a complete description of Nature. 
The cosmological observations of the last thirty years, for instance, have revealed that the SM constituents explain only a small share of the total energy budget of the Universe. In particular, the analysis of the microwave radiation background shows that baryons constitute only about 15\% of all matter~\cite{Planck:2018vyg}. The remaining part is accounted for by dark matter (DM), a substance of unknown nature which finds no description in the SM. Presently, the leading direct detection experiments have not yet found clear signals of DM scattering on nucleons or electrons, resulting in upper bounds on the direct detection cross sections~\cite{LUX:2016ggv,XENON:2018voc,PandaX-4T:2021bab}. Similarly, this far collider searches have not found any presence of DM particles in the produced states~\cite{ATLAS:2021enr}. 

Problems arise also within the known boundaries of the SM, where the measured properties of the Higgs boson seem to indicate the metastability of the electroweak (EW) vacuum. Furthermore, the apparent absence of new degrees of freedom at the EW scale questions the known mechanisms deemed responsible for the stabilisation of the Higgs boson mass scale, which seemingly remains insensitive to the large radiative contributions expected within most of the ultraviolet completions of the theory.  

This lack of signals gives encouragement to look for other avenues in the attempt to pinpoint the possible physics beyond the SM. In regard of this, one important possibility is provided by the recent discovery of gravitational waves (GW) by the LIGO experiment~\cite{LIGOScientific:2016aoc,LIGOScientific:2016sjg}, which has opened a new window on the dynamics taking place in the early stages of the Universe. Crucially, GW astronomy allows us to explore cosmological phase transitions related to the dynamics of EW symmetry breaking. Given the measured properties of the Higgs boson, the EW phase transition (PT) predicted by the SM is a smooth cross-over~\cite{Kajantie:1996mn,Aoki:1999fi} which does not generate observable GW signals. However, scalar extensions of the SM can easily admit first-order phase transitions (FOPT) resulting in a stochastic GW background~\cite{Hogan:1983ixn,Steinhardt:1981ct,Witten:1984rs} within the reach of future space-based GW observatories such as LISA~\cite{eLISA:2013xep,2017arXiv170200786A}, BBO~\cite{Corbin:2005ny,Crowder:2005nr}, Taiji~\cite{Hu:2017mde,Ruan:2018tsw}, TianQuin~\cite{TianQin:2015yph} or DECIGO~\cite{Seto:2001qf,Kawamura:2020pcg}. Because the LHC discovery confirmed the existence of scalar fields in Nature, it is worth to consider possible applications of such models within particle physics and cosmology. 

Following this line of reasoning, in the present paper we revisit the phenomenology of the Inert Doublet Model (IDM)~\cite{Deshpande:1977rw,Ma:2006km,Barbieri:2006dq,LopezHonorez:2006gr}, paying special attention to the reach of future GW experiments. The IDM is characterised by the presence of a second scalar doublet stabilised by a $\mathbb{Z}_2$ symmetry which -- crucially -- is \textit{not} spontaneously broken. Consequently, the new field is `inert': it cannot couple to the SM fermions via Yukawa interactions and it is forbidden from acquiring a non-vanishing vacuum expectation value (VEV) at zero temperature.  

In spite of these restrictions, the IDM can give rise to a rich phenomenology.\footnote{See for instance~\cite{Blasi:2022woz} for the impact of topological defects in two-step phase transitions.} For instance, owing to the unbroken $\mathbb{Z}_2$ symmetry, the lightest neutral component of the new doublet is necessarily stable and, therefore, constitutes a viable DM candidate. 
In regard of this, previous studies have shown that the IDM can account fully for the observed DM relic density in two regions of its parameter space. One possibility is provided by the Higgs resonance region, where the mass of DM is close to half of the Higgs boson mass. Alternatively, in the large-mass region, the new scalar components have a compressed mass spectrum~\cite{Goudelis:2013uca,Blinov:2015qva,Diaz:2015pyv}.

The DM and collider phenomenology of the IDM are well studied~\cite{Belanger:2015kga,Datta:2016nfz,Belyaev:2022wrn} (see e.g. Refs.~\cite{Belyaev:2016lok,Belyaev:2018ext} for a recent review). The new scalar interactions in the IDM can also stabilise the EW vacuum (e.g. \cite{Khan:2015ipa}). The phase diagrams of the IDM and the dependence of phase structure on the values of scalar quartic couplings were elucidated in Ref.~\cite{Ginzburg:2010wa}. While in the literature there are studies of the EW phase transitions within the IDM~\cite{Huang:2017rzf,Wang:2020wrk,Chowdhury:2011ga,Borah:2012pu,Gil:2012ya,  Blinov:2015vma, Fabian:2020hny}, we find that a full assessment, including a consistent treatment of bubble nucleation and the implied GW signal, is still missing. The possibility of realizing multi-step phase transitions~\cite{Fabian:2020hny}
, in particular, is intriguing because these processes can generate the multi-peak GW spectra considered first in Ref.~\cite{Morais:2019fnm}. This fascinating signature has been studied within the context of other two-Higgs-doublet models~\cite{Land:1992sm, Zarikas:1995qb, Aoki:2021oez}, where the restrictions of the IDM are lifted. Within the IDM, instead, the cross-correlation between DM and phase transition phenomenology was investigated in Ref.~\cite{Cline:2013bln, Fabian:2020hny}.

With the present paper, we intend to improve on existing analyses pertaining to cosmic phase transitions within the IDM. To this purpose, we analyse the parameter space allowed by the latest collider and DM searches in the attempt to map the available phase transition patterns, as well as the GW signals they produce. Although most commonly the EW phase transitions occur in a single step ($O \to h$), we find regions of the parameter space where two-step ($O \to H \to h$ or $O \to hH \to h$) and even three-step ($O \to H \to hH \to h$) transitions are realised. We pay particular attention to two- and three-step processes that involve multiple first-order phase transitions, which have the potential to generate a clear GW signature presenting multiple peaks in the spectrum. In our analysis we impose that the DM relic density produced by the IDM does not exceed the $3\sigma$ upper bound indicated by the latest Planck measurement \cite{Planck:2018vyg}. We also assess the impact of direct detection experiments, finding that they significantly constrain the parameter space yielding multi-step phase transitions with the exception of the Higgs resonance region. The latter will be completely probed in future monojet searches, which are therefore crucial to rule in or out the identified multi-step solutions.

In light of the new CDF determination of the $W$ boson mass~\cite{CDF:2022hxs}, we also present a separate analysis of the impact of the new measurement.
In agreement with previous results~\cite{Fan:2022dck}, we observe that adopting the CDF value strongly reduces the number of viable DM solutions for DM masses of order $\mathcal{O}(100)$ GeV or larger, especially for phase transition patterns involving at least one first-order transition.    

The paper is organised as follows. In Sec.~\ref{sec:idm}, after presenting the IDM framework, we summarise the results concerning the quantum and thermal corrections received by the scalar potential. In Sec.~\ref{sec:constrains}, instead, we define the parameter space probed by our analysis and detail the theoretical and experimental constraints that we apply. Sec.~\ref{sec:pt} discusses the evolution of the vacuum and the patterns of cosmic phase transition that we have identified. The results obtained are then scrutinized in Sec.~\ref{sec:dd_and_monojet}, where we analyse the impact of direct detection experiment and the power of future collider searches. The predicted GW signals are  given in Sec.~\ref{sec:gw}, whereas the implications of the new $W$ boson mass are detailed in Sec.~\ref{sec:wmass}. Finally, we conclude by summarising our findings in Sec.~\ref{sec:concl}.

\section{The inert doublet model}
\label{sec:idm}

\subsection{Tree-level potential}

The SM Higgs doublet $H_1$ and the inert doublet $H_2$ can be decomposed as 
\begin{equation}
    H_1 = \begin{pmatrix} G^+ \\ \frac{v + h + i G^0}{\sqrt{2}}  \end{pmatrix},
    \qquad
    H_2 = \begin{pmatrix} H^+ \\ \frac{H + i A}{\sqrt{2}}  \end{pmatrix},
\end{equation}
where $h$ is the SM Higgs boson, $v = 246.22~\mathrm{GeV}$ is the EW VEV and $G^+$ and $G^0$ are Goldstone bosons. The inert doublet comprises a charged scalar field $H^\pm$, and two neutral scalars, $H$ and $A$, with opposite CP-parities.

The tree-level potential of the model, 
\begin{align}
    \label{eq:V0}
  V &= -m_{1}^2 |H_1|^2  -m_{2}^2 |H_2|^2 + \lambda_1 |H_1|^4 + \lambda_2 |H_2|^4 +  \lambda_3 |H_1|^2 |H_2|^2  + \lambda_4 |H_1^\dagger H_2|^2 + \frac{\lambda_5}{2} \left[ (H_1^\dagger H_2)^2 + \mathrm{h.c.} \right],
\end{align}
respects a discrete $\mathbb{Z}_2$ symmetry under which $H_2$ is odd and all the SM fields are even. The symmetry thus ensures the stability of the lightest component of the inert doublet and forbids new Yukawa couplings between $H_2$ and the SM fermions, hence the epithet \emph{inert}.

The requirement that the tree-level potential be minimised at the EW vacuum leads to the following parametrisation 
\begin{align}
    m_{1}^2 &= \frac{m_h^2}{2}, & m_{2}^2 &= - m_H^2 + \lambda_{345} \frac{v^2}{2}, & \lambda_1 &= \frac{m_h^2}{2 v^2}, & \lambda_3 = \lambda_{345} +2 \frac{m_{H^\pm}^2 - m_H^2}{v^2},
    \nn \\
    \lambda_4 &= \frac{m_H^2 + m_A^2 - 2 m_{H^\pm}^2}{v^2}, & \lambda_5 &= \frac{m_H^2 - m_A^2}{v^2},
    \label{eq:parametrisation}
\end{align}
given in terms of the tree-level scalar mass matrix eigenvalues $m_h^2, m_H^2, m_A^2$ and $m_{H^\pm}^2$ ($m_{G^0}=m_{G^\pm}=0$ at tree-level in the EW vacuum).

The inert doublet self-coupling $\lambda_2$ does not affect DM phenomenology, but can influence the phase structure of the potential by inducing new minima at non-zero temperature. With the parametrisation in Eq.~\eqref{eq:parametrisation}, the model is completely specified by the quantities $\lambda_2$, $\lambda_{345}  = \lambda_3 + \lambda_4 + \lambda_5$, and the masses $m_H$, $m_{H^\pm}$, $m_A$, which we use as input parameters in our analysis. The lightest neutral components of $H_2$ is a viable DM candidate. In our analysis, this role is assigned to $H$, in effect choosing $\lambda_5 < 0$. Equivalently, $A$ could be the DM candidate, related to our case through the substitutions $\lambda_{345} \leftrightarrow \tilde{\lambda}_{345}= \lambda_3 + \lambda_4 - \lambda_5$ and $m_H\leftrightarrow m_A$.\footnote{In regard of this, notice that $\lambda_5\to-\lambda_5$ under the substitution $m_H\leftrightarrow m_A$ and that the quartic couplings determining the DM abundance via $hHH$ or $hAA$ interactions are, respectively, by $\lambda_{345}$ and $\tilde{\lambda}_{345}$.}

For the treatment of the phase transitions in Section~\ref{sec:pt}, we suppose that excursions in the field space occur only in the $(h, H)$ plane, while the remaining scalar degrees of freedom are prevented from acquiring a VEV at any temperature. Therefore, the terms in the tree-level potential relevant for this analysis are  
\begin{equation}
V_0(h, H) = -\frac{m_{1}^2}{2} h^2 +\frac{\lambda_1}{4}h^4 -\frac{m_{2}^2}{2} H^2 +\frac{\lambda_2}{4}H^4+\frac{\lambda_{345}}{4}h^2H^2.
\end{equation}

\subsection{Coleman-Weinberg correction to the potential}

The tree-level potential in Eq.~\eqref{eq:V0} receives important radiative contributions sourced by the one-loop $n$-point functions, re-summed in the Coleman-Weinberg correction~\cite{Coleman:1973jx}
\begin{equation}
    \label{eq:VCW}
    V_{\rm CW}(h, H) = \frac{1}{64 \pi^2} \sum_i n_i m_i^4 \left( \ln \frac{m_i^2}{\mu^2} - C_i \right),
\end{equation}
where $i = W, Z, t, h, H, G^0, A, G^\pm, H^\pm$ (as customary, we retain only the dominant fermion contribution given by the top quark), $\mu$ is the renormalisation scale (which we set to $\mu=v$) and $C_i$ are constants peculiar to the renormalisation scheme. The bosonic and fermionic contributions are weighted by the coefficients $n_i$ given by
$n_W = 6$, $n_Z = 3$, $n_t = -12$, $n_h = n_H=  n_{G^0} = n_A = 1$ and $n_{H^\pm} = n_{G^\pm} = 2$~\cite{Delaunay:2007wb}. After using dimensional regularisation with the $\overline{\rm MS}$ subtraction scheme, we have $C_i=\frac{3}{2}$ for scalars, fermions and longitudinal vector bosons, as well as $C_i=\frac{1}{2}$ for transverse vector bosons. The field-dependent masses $m_i^2$ in Eq.~\eqref{eq:VCW} include, respectively, the eigenvalues of the neutral and charged scalar mass matrices
\begin{equation}
\operatorname{M}_{S_1}^2 = \left(
\begin{array}{cc}
- m_{1}^2 + 3 \lambda_1 h^2 + \frac{ \lambda_{345}}{2} H^2		&  \lambda_{345} \, h \, H  \\
\lambda_{345} \, h \, H 		&		- m_{2}^2 + 3 \lambda_2 H^2 + \frac{ \lambda_{345}}{2} h^2
\end{array}
\right) , 
\end{equation}
\begin{equation}
\operatorname{M}_{S_2}^2 = \left(
\begin{array}{cc}
- m_{1}^2 + \lambda_1 h^2 + \frac{\tilde \lambda_{345}}{2} H^2		&  \lambda_5 \, h \, H  \\
\lambda_5 \, h \, H 		&		- m_{2}^2 + \lambda_2 H^2 + \frac{\tilde \lambda_{345}}{2} h^2
\end{array}
\right) , 
\end{equation}
\begin{equation}
\operatorname{M}_{\pm}^2 = \left(
\begin{array}{cc}
-m_{1} + \lambda_1 h^2 + \frac{\lambda_3}{2} H^2		&  \frac{\lambda_4 + \lambda_5}{2} h \, H   \\
\frac{\lambda_4 + \lambda_5}{2} h \, H 		&		-m_{2} + \lambda_2 H^2 + \frac{\lambda_3}{2} h^2
\end{array}
\right)  ,
\end{equation}
as well as the EW gauge bosons and top-quark contributions
\begin{equation}
m_W^2 = \frac{1}{4} g_2^2 (h^2 + H^2) , \qquad  m_Z^2 = \frac{1}{4} (g_1^2 + g_2^2) (h^2 + H^2) , \qquad
m_t^2 = \frac{1}{2} y_t^2 h^2,
\end{equation}
with $g_1$, $g_2$ and $y_t$ the $U(1)_Y$, $SU(2)_L$ and top-quark Yukawa couplings respectively.

Following previous analyses~\cite{Cline:1996mga, Cline:2011mm, Fabian:2020hny}, we compensate possible radiative shifts of the EW VEV and address the problematic Goldstone contributions with a set of counterterms specified in 
\begin{equation}
    \label{eq:VCT}
    V_{\rm CT}(h,H) = \delta m_h ^2 h^2 + \delta m_H^2 H^2 + \delta \lambda_1 h^4,
\end{equation}
where the coefficients are determined by the renormalisation conditions 
\begin{align}
    \frac{\pd V_{\rm CT}}{\pd h}\Big|_{\rm VEV} &= -\frac{\pd V_{\rm CW}}{\pd h}\Big|_{\rm VEV} ,
    \\
    \frac{\pd^2 V_{\rm CT}}{\pd h^2}\Big|_{\rm VEV} &= 
    -\left(\frac{\pd^2 V_{\rm CW}|_{G^0, G^\pm \equiv 0}}{\pd h^2}
    + \frac{1}{32 \pi^2}\sum_{G=G^0, G^\pm}\left(\frac{\pd m^2_G}{\pd h}\right)^2 \ln\left(\frac{m^2_{\rm IR}}{\mu^2}\right)
    \right)\Bigg{\vert}_{\rm VEV} ,
    \\
    \frac{\pd^2 V_{\rm CT}}{\pd H^2}\Big|_{\rm VEV} &= 
    -\left(\frac{\pd^2 V_{\rm CW}|_{G^0, G^\pm \equiv 0}}{\pd H^2}
    + \frac{1}{32 \pi^2}\sum_{G=G^0, G^\pm}\left(\frac{\pd m^2_G}{\pd H}\right)^2 \ln\left(\frac{m^2_{\rm IR}}{\mu^2}\right)
    \right)\Bigg{\vert}_{\rm VEV} ,
\end{align}
where VEV here corresponds to the zero-temperature vacuum $(h,H)=(v,0)$ and $m_{\rm IR}$ is the infrared cutoff used for the regularisation of Goldstone contributions, which we set to the Higgs boson mass, $m_{\rm IR} = m_h$~\cite{Cline:2011mm}.\footnote{Further improvements can be obtained by using the effective potential in a dimensionally reduced field theory \cite{Niemi:2021qvp,Schicho:2021gca,Schicho:2022wty,Croon:2020cgk}.}

\subsection{Finite temperature effects}
At finite temperature, thermal corrections result in a further contribution~\cite{PhysRevD.9.3320}, 
\begin{equation}
    \label{eq:VT}
    V_{\rm T} (h, H, T) = \frac{T^4}{2\pi} 
    \left[
    \sum_i n_i^{\rm B} J_{\rm B}\left(\frac{m_i^2}{T^2}\right)
    +
    \sum_i n_i^{\rm F} J_{\rm F}\left(\frac{m_i^2}{T^2}\right)
    \right],
\end{equation}
to the scalar potential. The two sums are over the boson and fermion degrees of freedom, respectively 
and the corresponding thermal functions~\cite{PhysRevD.45.2685} are
\begin{equation}
    J_{\rm B/F}(x) = \int\limits_0^\infty \td t\, t^2 \ln\left(1 \mp e^{-\sqrt{t^2+x}}\right)\, .
\end{equation}

A consistent treatment of thermal corrections also requires the resummation of the leading self-energy daisy diagrams, which shifts the field-dependent masses 
\begin{equation}
m_{1}^2(T) = m_{1}^2 - c_1 T^2 
\,, \qquad 
m_{2}^2(T) = m_{2}^2 - c_2 T^2 , 
\end{equation}
by a thermal contribution quantified in the coefficients~\cite{RevModPhys.53.43, PhysRevD.45.4695} 
\begin{align}
c_1 &=  \frac{1}{16} (g_1^2 + 3 g_2^2) + \frac{1}{4} y_t^2  +  \frac{6 \lambda_1 + 2 \lambda_3 + \lambda_4}{12} ,  \\
c_2 &=  \frac{1}{16} (g_1^2 + 3 g_2^2) +  \frac{6 \lambda_2 + 2 \lambda_3 + \lambda_4}{12}.  
\end{align}

Thermal corrections also result in Debye masses for the longitudinal components of gauge bosons~\cite{Bernon:2017jgv}, given by
\begin{align}
    m^2_{W_L} &= \frac{h^2 + H^2}{4} g_2^2 + 2  g_2^2 \, T^2,
    \\
    m^2_{Z_L} &= \frac{h^2 + H^2}{8} \left(g_1^2 + g_2^2\right) + \left(g_1^2 + g_2^2\right) T^2 +\Delta ,
    \\
    m^2_{\gamma_L} &= \frac{h^2 + H^2}{8} \left(g_1^2 + g_2^2\right) + \left(g_1^2 + g_2^2\right) T^2  -\Delta ,
\end{align}
with
\begin{equation}
    \Delta^2 = \frac{(h^2 + H^2 +8 T^2)^2}{64}\left(g_1^2 + g_2^2\right)^2 - g_1^2\, g_2^2 \, T^2
    \left(h^2 + H^2 +4T^2\right).
\end{equation}
In our analysis we use the above thermal masses when computing the CW and the finite temperature corrections to the tree-level potential. The full thermally-corrected effective potential is thus
\begin{equation}
\label{eq:Veff}
V_{\rm eff}(h, H, T) =  V_0(h, H) + V_{\rm CW}(h, H, T) + V_{\rm CT}(h, H) +V_{\rm T}(h, H, T).    
\end{equation}

\section{Parameter space and considered constraints}
\label{sec:constrains}

With the full expression of the scalar potential at hand, we briefly review the constraints applied in the forthcoming analysis and define the explored region of IDM the parameter space.

\subsection{Considered parameter ranges}

In our analysis, we scan the parameter space shown in Tab.~\ref{tab:ranges}. We then use the {\tt CosmoTransitions} package~\cite{Wainwright:2011kj} to obtain, for each point selected, the temperature-dependent phase structure of the scalar potential and to assess the nature of the corresponding phase transitions.


\begin{table}[h]
    \centering
    \begin{tabular}{|c|c|}
    \hline
        Parameter & Range \\
    \hline
        $m_H$ &  $[10, 1000]$ GeV \\
        $m_A$ &  $[10, 1000]$ GeV \\
        $m_{H^+}$ & $ [10, 1000]$ GeV \\
        $\lambda_2 $ & $[0, \frac{4\pi}{3}]$ \\
        $\lambda_{345}$ & $ [-1.47, 4\pi]$\\
        \hline
    \end{tabular}
    \caption{The parameter ranges used in our scan. We selected only configurations with $m_H<m_A$, since $H$ is our DM candidate. The lower bound on $\lambda_{345}$ is imposed by the stability of the potential~\cite{Belyaev:2016lok}.}
    \label{tab:ranges}
\end{table}

The obtained points are then selected according to the bounds discussed below.

\subsection{Theoretical constraints}
\label{sec:constr:th}
A first requirement is the stability of the scalar potential, which guarantees that minima appear at finite field values. For the IDM, the potential is  bounded from below if the following conditions are satisfied:
\begin{equation}
    \lambda_1 > 0, 
    \quad 
    \lambda_3 + 2 \sqrt{\lambda_1 \lambda_2} > 0,
    \quad
    \lambda_3 + \lambda_4 - |\lambda_5| + 2 \sqrt{\lambda_1 \lambda_2} > 0.
\end{equation}

A charge-breaking vacuum is avoided by $\lambda_4 - |\lambda_5| < 0$, which always holds if $H^\pm$ is heavier than the DM candidate $H$
~\cite{Ginzburg:2010wa}.

Perturbative unitarity requires that the combinations of couplings $e_i$ from the eigenvalues of the two-to-two scattering matrix be bounded: $|e_i| < 8 \pi$. From the full $22 \times 22$ $S$-matrix \cite{Arhrib:2012ia}, we have \cite{Belyaev:2016lok} $e_{1,2}= \lambda_3 \pm \lambda_4, \ \ e_{3,4}=\lambda_3 \pm \lambda_5, \ \ \
e_{5,6}= \lambda_3 + 2\lambda_4 \pm 3\lambda_5, \ \ \  e_{7,8} = -\lambda_1-\lambda_2 \pm \sqrt{(\lambda_1-\lambda_2)^2+\lambda_4^2}$, $e_{9,10} = -3\lambda_1 - 3\lambda_2 \pm \sqrt{9(\lambda_1 -\lambda_2)^2 + (2\lambda_3+\lambda_4)^2}, \ \ \ 
 e_{11,12} = -\lambda_1 - \lambda_2 \pm \sqrt{(\lambda_1 - \lambda_2)^2 + \lambda_5^2}
$.
 The strongest constraints are given by $|\lambda_2| < 4 \pi/3$ and $|\lambda_{345}| < 4 \pi$.

\subsection{Experimental constraints}
\label{sec:constr:exp}

The decay widths of the $Z$ and $W$ bosons measured at LEP with high precision preclude decays of these particles into the new states. Therefore, we require that the masses of the inert doublet components satisfy~\cite{Cao:2007rm}
\begin{equation}
    m_H + m_{H^\pm} > m_W, \quad m_A + m_{H^\pm} > m_W, \quad m_H + m_A > m_Z, \quad 2 m_{H^\pm} > m_Z.
\end{equation}
LEP searches for new neutral final states further exclude a range of masses~\cite{Lundstrom:2008ai}, thereby forcing
\begin{equation}
    m_H > 80~\mathrm{GeV}, \quad m_A > 100~\mathrm{GeV} \quad\text{or} \quad m_A - m_H < 8~\mathrm{GeV},
\end{equation}
in addition to
\begin{equation}
    m_{H^\pm} > 70~\mathrm{GeV}
\end{equation}
due to searches for charged scalar pair production~\cite{Pierce:2007ut}. 

Similarly, if $m_H < m_h/2$, the Higgs boson can decay into DM with a partial width of
\begin{equation}
    \Gamma_{h \to HH} = \frac{\lambda_{345}^2 v^2}{32 \pi m_h} \sqrt{1 - \frac{4 m_H^2}{m_h^2}}
\end{equation}
which is constrained by measurements of the Higgs boson invisible width. The current values provided by the ATLAS and CMS experiments~\cite{ATLAS-CONF-2018-031,CMS:2016dhk} on the invisible branching ratio $\text{BR}_\text{inv} = \Gamma_{h \to HH} / (\Gamma_{h \to \text{SM}} + \Gamma_{h \to HH})$ are $\text{BR}_\text{inv} < 0.23-0.36$. In the following, we will use the conservative limit $\text{BR}_\text{inv} < 0.23$.

Collider analyses also provide constraints on the electroweak precision observables (EWPO), sensitive to new radiative contributions in the electroweak sector. The EWPO are usually expressed via the Peskin-Takeuchi parameters $S$, $T$ and $U$~\cite{Peskin:1990zt,Peskin:1991sw}, determined through a joint fit of the precision observables and SM predictions. Disregarding for the moment the new determination of the $W$ boson mass by the CDF collaboration~\cite{CDF:2022hxs} (discussed in Sec.~\ref{sec:wmass}), the EWPO fit within the SM alone~\cite{Lu:2022bgw} gives the results presented in Tab.~\ref{tab:STU}.

\begin{table}[h]
    \centering
    \begin{tabular}{|c|c|c|}
    \hline
    Parameter & Result & Correlation \\
    \hline
         $S$ &  $0.06 \pm 0.10$ &   $0.90$ $(T)$, $-0.57$ $(U)$ \\
         $T$ & $0.11 \pm 0.12$ &  $-0.82$ $(U)$ \\
         $U$ & $-0.02 \pm 0.09$ & \\
         \hline
    \end{tabular}
    \caption{Peskin-Takeuchi parameters~\cite{Peskin:1990zt,Peskin:1991sw} as determined by the electroweak precision observables~\cite{Lu:2022bgw} prior to the new $W$ boson mass determination by the CDF collaboration~\cite{CDF:2022hxs}.}
    \label{tab:STU}
\end{table}

The IDM contributions to the $S$, $T$ and $U$ parameters, which add to the SM result, are given 
by~\cite{Grimus:2008nb}

\begin{equation}
T = \frac{1}{32\pi^2\alpha v^2}\left[F(m_{H^\pm}^2,m_{A}^2) + F(m_{H^\pm}^2,m_{H}^2) - F(m_{A}^2,m_{H}^2)\right],
\end{equation}
where  $F(x,y) = \frac{x+y}{2}-\frac{xy}{x-y}\ln (x/y)$ for $x\neq y$ and $f_c(x,x) = 0$ and

\begin{equation}
    S=\frac{4\sin^2\theta_W}{\alpha}\frac{g^2}{384\pi^2}\left[ (\cos^2\theta_W-\sin^2\theta_W)^2G(m_{H^\pm}^2,m_{H^\pm}^2,m_Z^2)+G(m_H^2,m_A^2,m_Z^2)+\ln\frac{m_A^2}{m_{H^\pm}^2}+\ln\frac{m_H^2}{m_{H^\pm}^2}\right],
\end{equation}

\begin{equation}
\begin{split}
U &=  \frac{4\sin^2\theta_W}{\alpha}\frac{g^2}{384\pi^2}\left[ -(\cos^2\theta_W-\sin^2\theta_W)^2G(m_{H^\pm}^2,m_{H^\pm}^2,m_Z^2)+G(m_{H^\pm}^2,m_A^2,m_W^2)+G(m_{H^\pm}^2,m_H^2,m_W^2) \right.
\\
& \left. ~~~-G(m_H^2,m_A^2,m_Z^2)\right],
\end{split}
\end{equation}
where $G(I,J,Q)\equiv-\frac{16}{3}+\frac{5(I+J)}{Q}-\frac{2(I-J)^2}{Q^2}+\frac{3}{Q}[\frac{I^2+J^2}{I-J}\frac{I^2-J^2}{Q}+\frac{(I-J)^3}{3Q^2}]\ln\frac{I}{J}+\frac{r}{Q^3}f(t,r)$, with $t\equiv I+J-Q$ and $r\equiv Q^2-2Q(I+J)+(I-J)^2$ and where 
\begin{equation}
    f(t,r)\equiv \begin{cases}
      \sqrt{r}\ln\vert\frac{t-\sqrt{r}}{t+\sqrt{r}}\vert & \text{if~} r>0\\
      0 & \text{if~} r=0\\
      2\sqrt{-r}\arctan\frac{\sqrt{-r}}{t} & \text{if~} r<0.
    \end{cases}  
\end{equation}

Notice that new EWPO contributions vanish in the limit of degenerate masses, so these observables tend to discourage hierarchical mass spectra. For the purpose of constraining the IDM parameter space, we require that the total values of $S$, $T$ and $U$ remain within the $95\%$ joint confidence level.

Finally, the properties of our DM candidate are constrained by the latest Planck measurements, which gives the corresponding relic density as $\Omega_c h^2 = 0.120 \pm 0.001$~\cite{Planck:2018vyg}. In our analysis we impose the $3 \sigma$ \emph{upper} bound indicated by the data, although we allow for the possibility that the inert doublet yield only a subdominant DM component. For the computation of the relic abundance we rely on the {\tt micrOMEGAs} code~\cite{Belanger:2020gnr}.

\section{Phase transitions}
\label{sec:pt}


In the early Universe, thermal evolution of the IDM effective potential can cause cosmic PTs and generate GWs.
At high temperatures thermal corrections dominate the potential and force the fields to vanishing VEVs at the origin of the field space. As the temperature lowers with the expansion of the Universe, additional minima appear in the scalar potential at finite field values. At the so-called critical temperature $T_c$, two minima are degenerate. Then at a lower temperature, PTs may then occur via tunneling across the potential barrier (between these two minima) sourced by  thermal effects.
The rate of these processes (the bubble nucleation rate due to false vacuum decay) per unit volume and time is given by~\cite{Linde:1980tt, Linde:1981zj}
\begin{equation}
    \Gamma(T) \simeq T^4 \left(\frac{S}{2\pi T} \right)^{3/2} e^{-S/T},
\end{equation}
where $S\equiv S(h,H,T)$ is the 3-dimensional Euclidean action computed for an $O(3)$-symmetric critical bubble (i.e. for the ``bounce'' solution of the equations of motion).\footnote{In the case of strong supercooling, one should also consider usual quantum tunnelling~\cite{Ellis:2018mja, Fujikura:2019oyi}.} The PT begins at the nucleation temperature $T_n$. This is defined as the temperature at which one bubble nucleates per Hubble volume per Hubble time 
\begin{equation}
    \Gamma(T_n) H_n^{-4} \sim O(1),
    \label{eq:end_PT}
\end{equation}
where $H_n\equiv H(T_n)$ is the value of the Hubble parameter at $T_n$. For FOPTs close to the EW scale, the condition above simply amounts to the criterion $S/T_n \simeq 140$.

The PT strength is quantified in the parameter~\cite{Espinosa:2010hh}
\begin{equation}
\label{eq:alpha}
\alpha = \frac{\Delta V_{\rm eff} - \frac{T}{4}\Delta\frac{\partial V_{\rm eff}}{\partial T}}{\rho_{\rm rad}}\Big|_{T=T_*},    
\end{equation}
where $\Delta V_{\rm eff} \equiv V_{\rm eff} \big |_{\text{false vacuum}} - V_{\rm eff} \big |_{\text{true vacuum}}$ (and similarly for $\Delta \frac{\partial V_{\rm eff}}{\partial T}$), the thermally corrected effective potential $V_{\rm eff}$ is defined in Eq.~\eqref{eq:Veff} and the radiation energy density is $\rho_{\rm rad} = \frac{\pi^2}{30} g T^4$, with $g$ the effective number of relativistic degrees of freedom at the temperature $T$. The subscript $*$ denotes parameters evaluated at the time when GWs are produced. In this paper we assume no significant reheating and can thus safely consider $T_*\simeq T_n$~\cite{Caprini:2015zlo}.

The PT inverse time-duration, on the other hand, is quantified in the parameter $\beta$~\cite{Grojean:2006bp} with
\begin{equation}
    \label{eq:beta}
    \frac{\beta}{H_n} = T_n\frac{d(S/T)}{dT}\Big |_{T_n},
\end{equation}
where the adiabatic time-temperature relation $\frac{dT}{dt} = -H T$ was used. The inverse of $\beta/H_n$ can also be used to estimate the number of bubbles per Hubble volume obtained at the end of the phase transition.\footnote{Indeed, the number of bubbles per Hubble volume during the phase transition is $\Gamma H^{-3}\beta^{-1}$. Therefore, when the PT completes at $T=T_n$, Eq.~\eqref{eq:end_PT} indicates that $H_n/\beta$ bubbles per Hubble volume were formed~\cite{Grojean:2006bp}.}

As we will show below, the use of the nucleation temperature in our computation of the GW signal (in place of the percolation temperature) is justified by the fact that the ratio $\beta/H_n$ is large almost everywhere in the considered parameter space.

\subsection{Phase transition patterns}

\begin{figure}[h!]
\begin{center}
\includegraphics[scale=1.5,trim = 3cm 21.2cm 14cm 6.cm, clip]{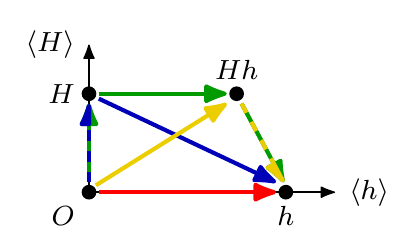}
\textit{\caption{Schematic representation of the possible phases and PT patterns supported by the IDM scalar potential. The high-temperature minimum of the potential, where $\langle h \rangle=\langle H \rangle=0$, is denoted with $O$. The phase $h$ is characterised by $\langle h \rangle\neq 0$ and $\langle H \rangle =0$, and includes the EW vacuum. The configuration where $\langle H \rangle\neq 0$ but $\langle h \rangle =0$ is denoted with $H$, while a phase with $\langle h \rangle, \langle H \rangle \neq0$ is indicated with $Hh$. The arrows show the different PT sequences identified in our analysis. \label{fig:phases:transs}
} }
\end{center}
\end{figure}

The PT patterns found in our analysis are summarised in Fig.~\ref{fig:phases:transs}, which schematically shows the sequences of transitions that connect the high-temperature minimum of the IDM potential, $O$, to the EW vacuum phase $h$. The red arrow indicates  one-step PTs $O \to h$, which directly connect the two minima. The blue arrows characterise two-step PTs $O \to H \to h$, in which the EW vacuum is reached after a transient phase, $H$, where only the inert doublet neutral component acquires a VEV. Similarly, the yellow arrows denote two-step PT patterns $O \to Hh \to h$ going through a different transient phase, $Hh$, in which both $h$ and $H$ acquire non-vanishing thermal VEVs. In our scan we have also identified three-step PTs $O \to H \to Hh \to h$ as indicated by the green arrows.

\begin{figure}[h!]
\begin{center}
\includegraphics[width=.99\linewidth,trim = 0cm 0cm 0cm 0cm,clip]{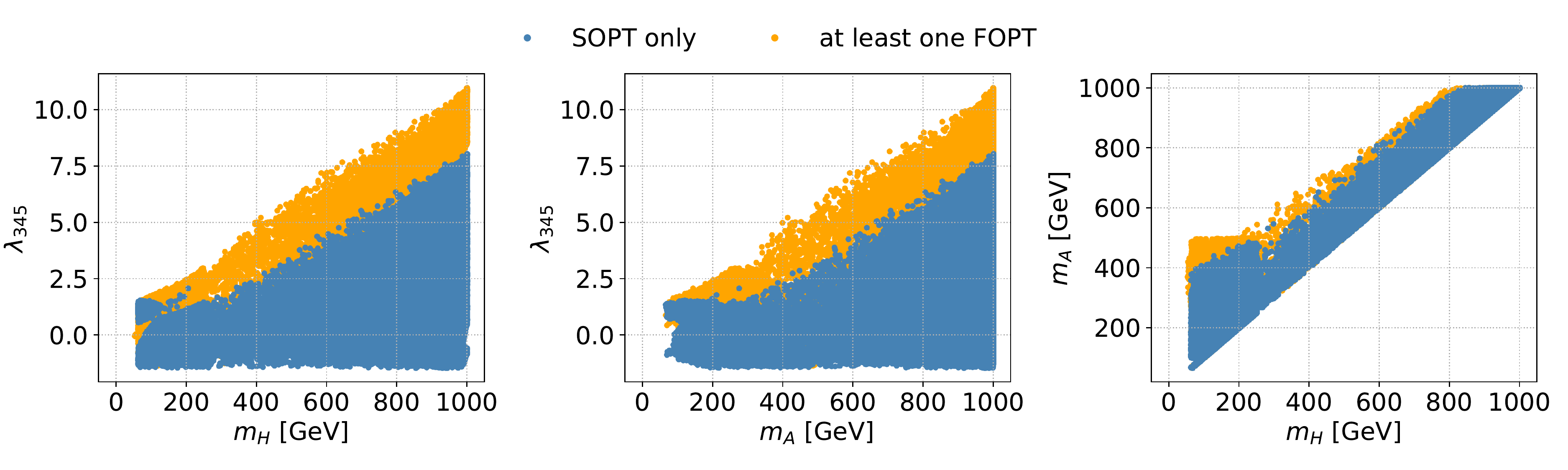}
\textit{\caption{Projections of the IDM parameter space on the planes spanned by $m_H$, $m_A$  and $\lambda_{345}$. Each panel shows the regions where the EW vacuum is achieved purely through PTs of the second order (blue) or PTs that involve at least one first-order process (orange). Points leading to DM overabundance are not shown.}\label{fig:SOPT_FOPT}}
\end{center}
\end{figure}

In the rest of this paper, we only consider strong FOPTs with $v_c/T_c \geq 1$, where $v_c \equiv\sqrt{\langle H\rangle^2_c+\langle h\rangle^2_c}$ is the VEV of the true vacuum at the critical temperature. This is motivated by the fact that for small $v_c/T_c$, perturbative analysis does not allow to clearly determine if a PT is of the first kind or not~\cite{Kajantie:1996mn}.


In Fig.~\ref{fig:SOPT_FOPT} we show the regions of the IDM parameter space where the identified PTs occur. In particular, the orange region indicates that the EW vacuum is reached by using at least one FOPT, thereby sourcing a potentially detectable GW signal. Conversely, the blue region shows transitions or sequences of transitions that involve exclusively second-order phase transitions (SOPTs) in each step. In these projections of the full parameter space, the blue region is contained within the orange one.

\begin{figure}[h!]
\begin{center}
\includegraphics[width=.99\linewidth,trim = 0cm 0cm 0cm 0cm, clip]{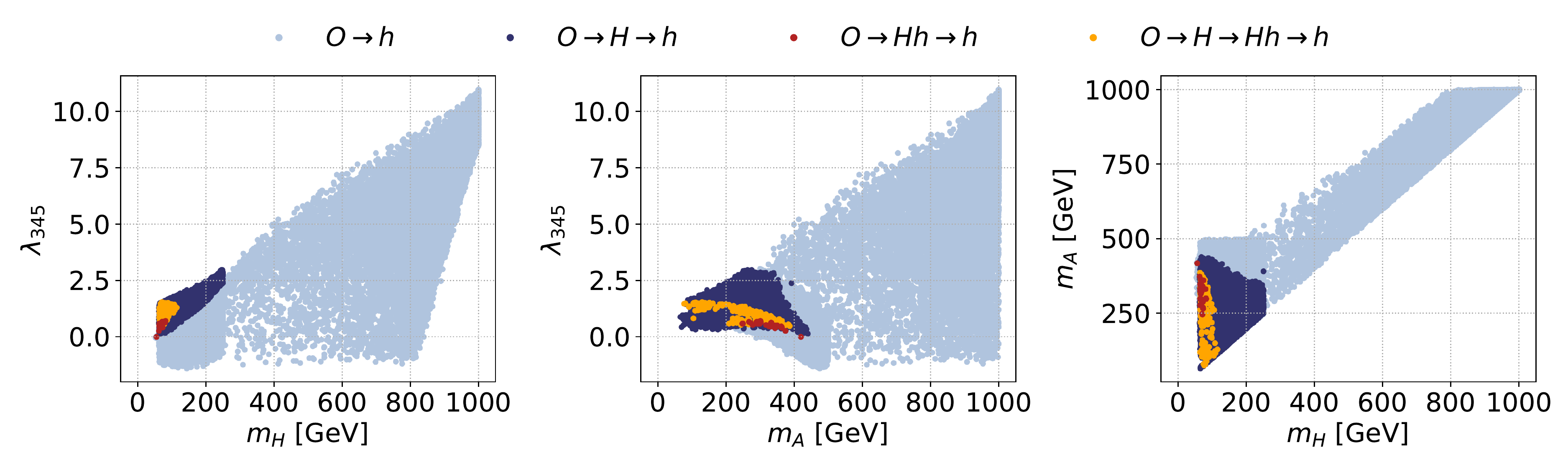}
\textit{\caption{Projections of the IDM parameter space on the planes spanned by $m_H$, $m_A$  and $\lambda_{345}$. Each panel shows the regions yielding a one-step PT $O\rightarrow h$ (light blue), a two-step PT $O\rightarrow H\rightarrow h$ (dark blue) and $O\rightarrow Hh \rightarrow h$ (red), as well as a three-step PT $O\rightarrow H\rightarrow Hh\rightarrow h$ (orange). For all these transitions, we require at least one FOPT. Points leading to DM overabundance are not shown.
}\label{fig:FOPT}}
\end{center}
\end{figure}

The sequences of PTs involving at least one FOPT step are presented in isolation in Fig.~\ref{fig:FOPT}. As we can see, most of the covered parameter space gives rise to one-step $O\rightarrow h$, whereas multi-step PTs only occur in a limited region roughly bounded by $0\lesssim \lambda_{345} \lesssim 3$, $m_H \lesssim  250$ GeV and $m_A, m_{H^+} < 500$ GeV, which we scan with greater accuracy. In particular, we find that three-step PTs require $\lambda_{345} \lesssim 1.5$, while two-step PTs via a transient $Hh$ phase are allowed only for $\lambda_{345}\lesssim 0.8$. 


\subsection{One-step phase transitions}

\begin{figure}[h!]
\begin{center}
\includegraphics[width=.99\linewidth,trim = 0cm 0cm 0cm 0cm, clip]{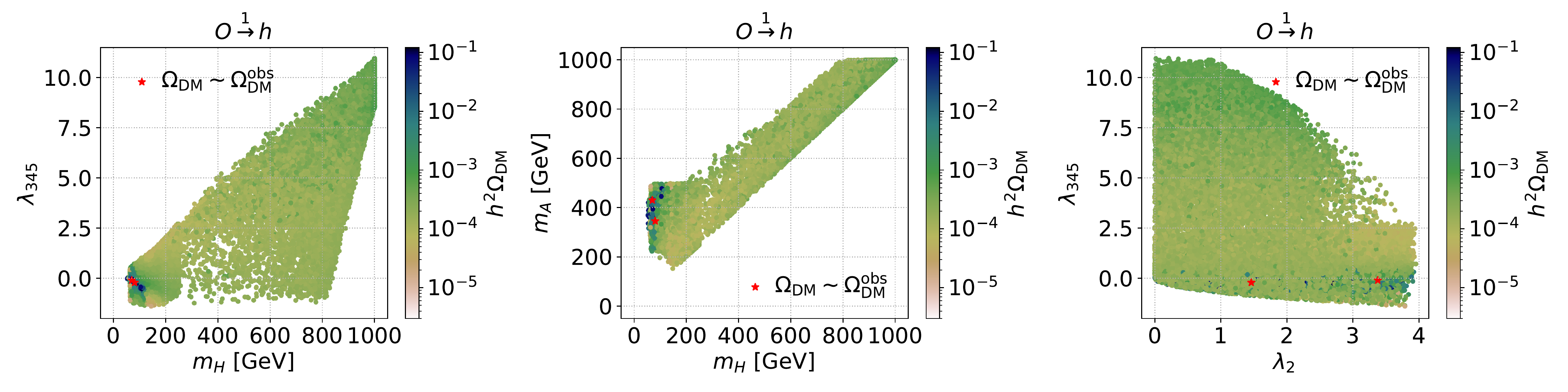}\\
\includegraphics[width=.99\linewidth,trim = 0cm 0cm 0cm 0cm, clip]{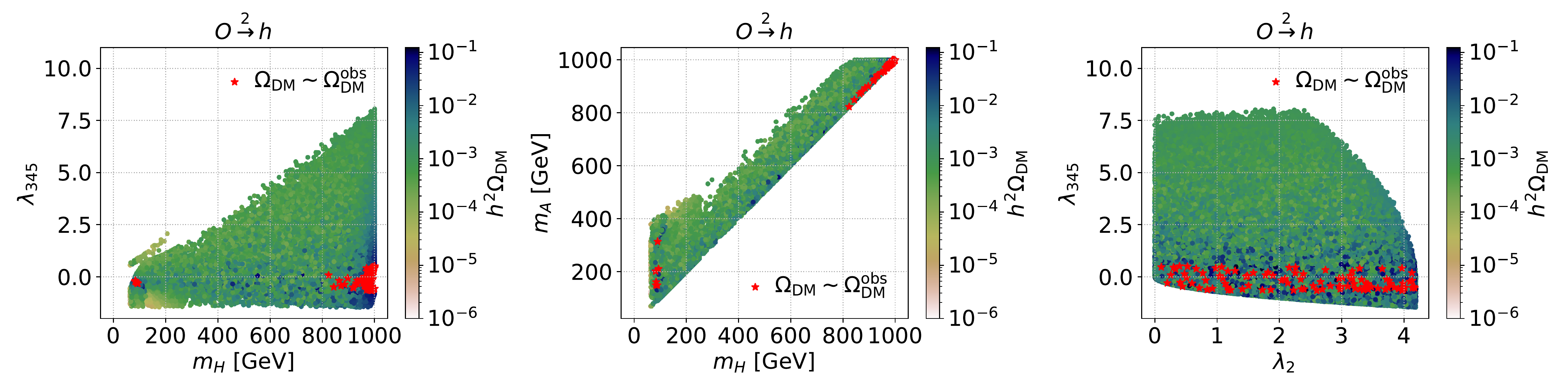}
\textit{\caption{Regions in the IDM parameter space yielding one-step PT $O\rightarrow h$. The three panels at the top illustrate the case of first-order PTs, while the bottom three panels show second-order PTs. The colour code indicates the obtained DM abundance, which agrees with the measured $3\sigma$ range in correspondence of red stars. Points leading to DM overabundance are not shown.}\label{fig:input_param_one_step}}
\end{center}
\end{figure}

Focusing on the identified one-step PTs, we show in Fig.~\ref{fig:input_param_one_step} the regions of the parameter space yielding a $O\rightarrow h$ transition. 

The panels in the top row show the case of FOPTs, whereas SOPTs are analysed in the bottom row. In all the plots, the colour code indicates the obtained DM relic density and red stars indicate points that lead to a relic density within the experimental 3$\sigma$ confidence interval.

As we can see, the considered values of scalar masses do not particularly favour either type of transition. Both FOPTs and SOPTs are furthermore uniformly distributed along the whole span of the analysed mass ranges. Comparing the plots in the first column of the figure shows that FOPTs generally require larger values of $\lambda_{345}$. For $m_H \gtrsim 800$ GeV, however, we notice that lower values of this coupling result exclusively in SOPTs. Close to the very end of the considered DM mass range, FOPTs thus require  $\lambda_{345}\gtrsim 8$. The value $\lambda_{345}\simeq 8$ also constitutes an upper bound obeyed by the SOPTs in the considered parameter space. As for the DM relic abundance, we see that FOPTs and SOPTs respectively prefer the small and large end of the investigated DM mass range. In either case, these solutions select small interval of $\lambda_{345}$ values centered on $\lambda_{345}\simeq 0$.

\subsection{Two-step phase transitions}

We have identified two patterns of two-step PT, $O\rightarrow H\rightarrow h$ and $O\rightarrow Hh\rightarrow h$, which have a different transient phase. Such processes are particularly relevant for the phenomenological exploration of the framework, as a sequence involving two FOPTs could result in a peculiar GW signal characterised by a double peak in the frequency spectrum.  

\begin{figure}[h!]
\begin{center}
\includegraphics[width=.99\linewidth,trim = 0cm 0cm 0cm 0cm, clip]{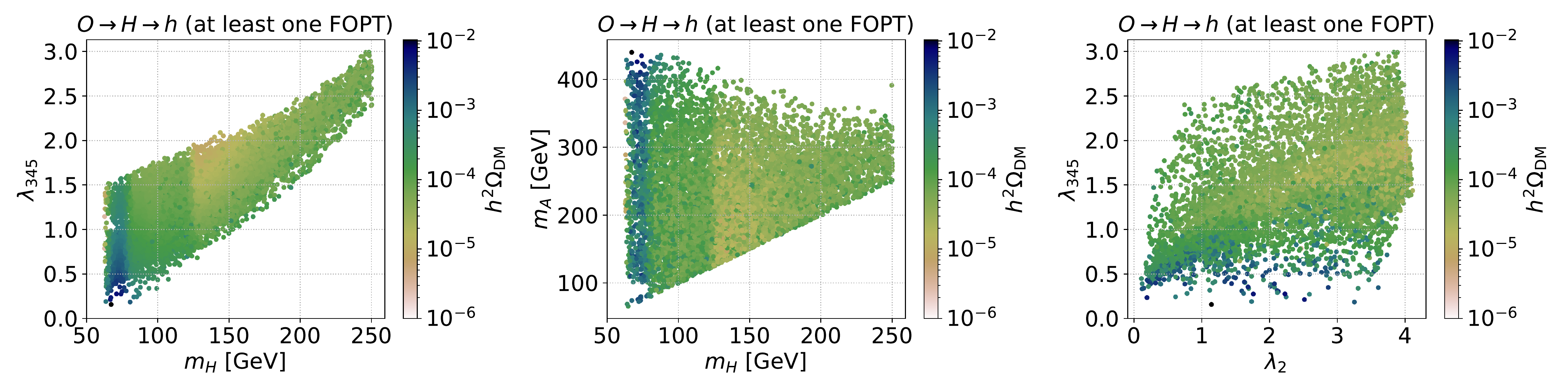}
\textit{\caption{Regions in the IDM parameter space yielding two-step PT $O\rightarrow H\rightarrow h$ involving at least one FOPT. The colour code indicates the obtained DM abundance, which for these solutions always falls below the $3\sigma$ range found by the Planck measurements. Points leading to DM overabundance are not shown.}\label{fig:input_param_osh}}
\end{center}
\end{figure}

Focusing for the moment on the sequence $O\rightarrow H\rightarrow h$, we notice that almost always these processes involve at least one FOPT step. The occurrence of this sequence in parameter space is shown in Fig.~\ref{fig:input_param_osh}. In the left panel, the linear trend between $\lambda_{345}$ and $m_H$ correlates with the produced dark matter abundance for $m_H\lesssim 200$ GeV, but it remains below the observed $3\sigma$ range even in the favoured low energy region. The middle panel shows that the DM abundance is relatively insensitive to the value of $m_A$ for a given DM mass and that, for large values of $m_H$, these solutions require a mild hierarchy between the new neutral states. The last plot, instead, shows that the obtained DM abundance is relatively insensitive to the strength of DM self-interactions regulated by $\lambda_2$. We stress that such solutions only yield an under-abundant DM relic density.

\begin{figure}[h!]
\begin{center}
\includegraphics[width=.99\linewidth,trim = 0cm 0cm 0cm 0cm, clip]{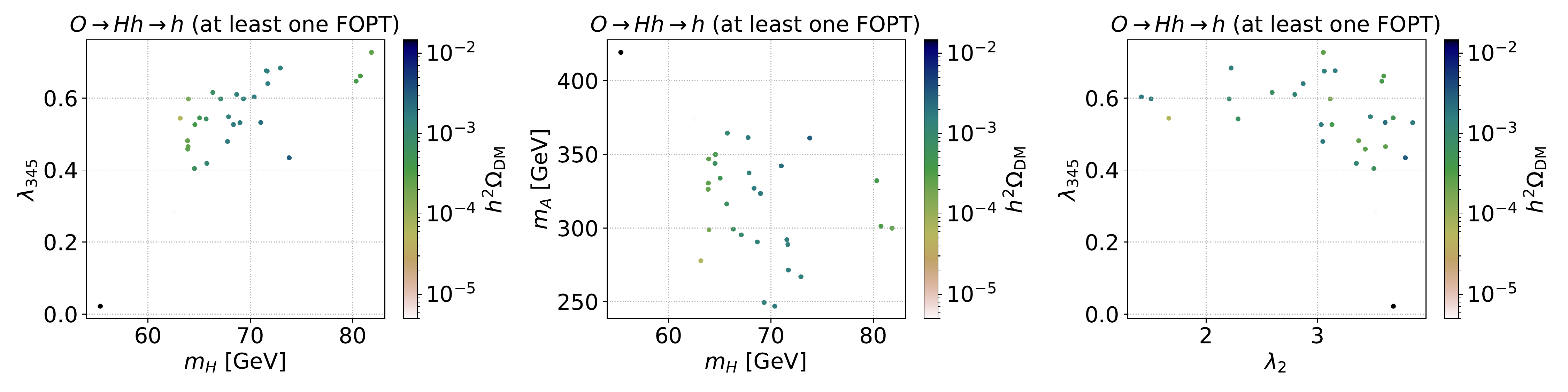}
\textit{\caption{Regions in the IDM parameter space yielding the two-step PT $O\rightarrow Hh\rightarrow h$ involving at least one FOPT. The colour code indicates the obtained DM abundance, which for these solutions always falls below the $3\sigma$ range found by the Planck measurements. Points leading to DM overabundance are not shown.}\label{fig:input_param_oshh}}
\end{center}
\end{figure}

Similarly to the previous case, the PT pattern $O\rightarrow Hh\rightarrow h$  predominantly has at least one FOPT. The values of the IDM parameters supporting this kind of transitions are shown in Fig.~\ref{fig:input_param_oshh}. As we can see, these solutions are rarer than those proceeding through a transient $H$ phase and select a narrow region in the IDM parameter space mostly characterised by small but positive values of $\lambda_{345}$ and a mild mass hierarchy of the neutral scalar states.

\subsection{Three-step phase transitions}

\begin{figure}[h!]
\begin{center}
\includegraphics[width=.99\linewidth,trim = 0cm 0cm 0cm 0cm, clip]{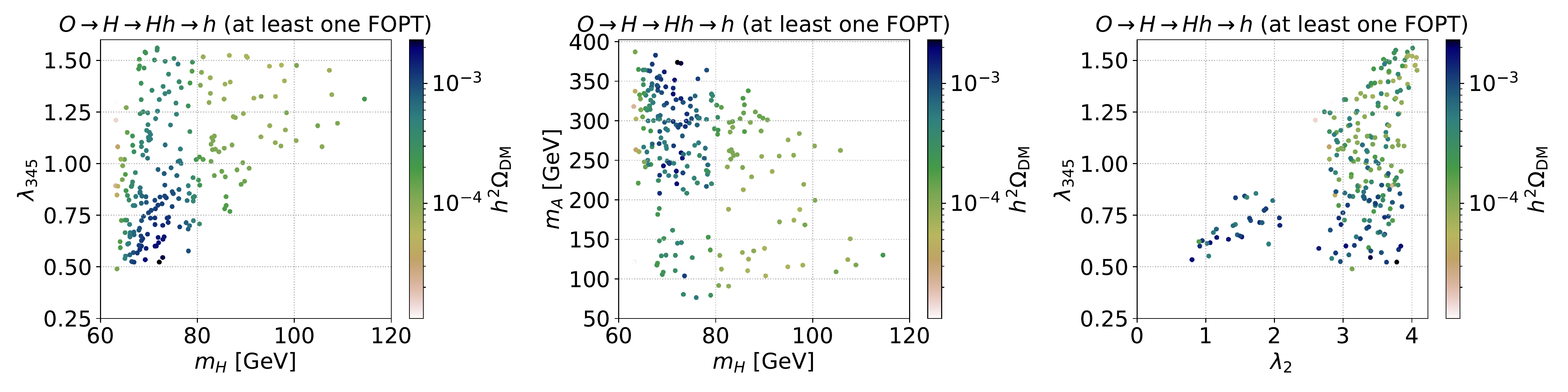}
\includegraphics[width=.99\linewidth,trim = 0cm 0cm 0cm 0cm, clip]{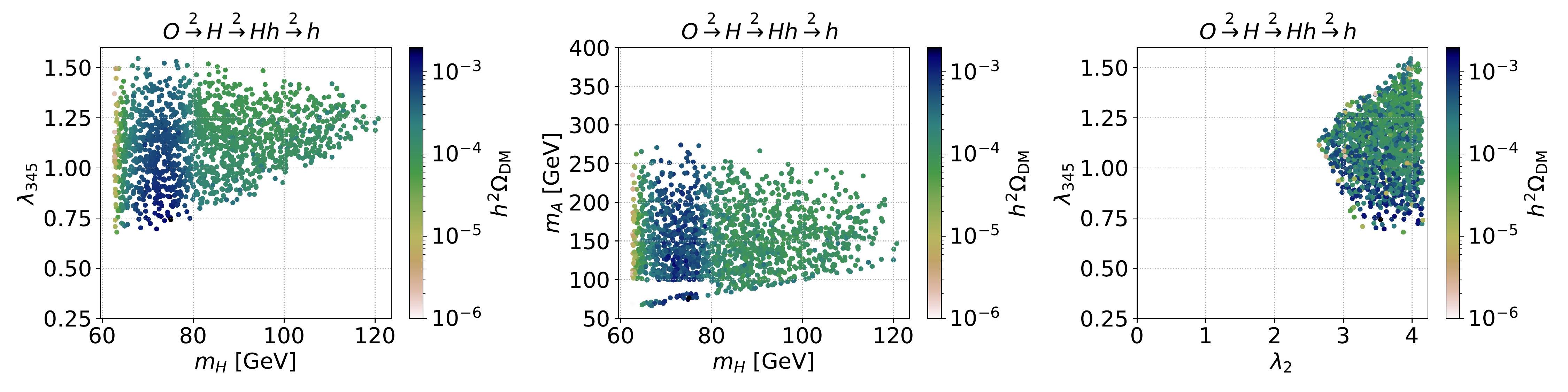}
\textit{\caption{Regions in the IDM parameter space yielding three-step PTs $O\rightarrow H\rightarrow Hh \rightarrow h$. The top row shows transitions where at least one of the steps is of the first order. The solutions shown in the bottom row, instead, have only SOPTs. In all the plots, the colour code indicates the obtained DM relic density. Points leading to DM overabundance are not shown.}\label{fig:input_param_osshh}}
\end{center}
\end{figure}

Three-step PTs generally occur for DM masses in a interval roughly given by $m_h/2 \lesssim m_H \lesssim m_h$ and require moderate values of the $\lambda_{345}$ coupling. Fig.~\ref{fig:input_param_osshh} shows the processes of this kind that we identified in our analysis, putting in evidence the IDM parameter space where at least one FOPT is involved in the transition chain. As we can see from the plots in the first row, such solutions requires a mild hierarchy in the neutral scalar sector that reduces with increasing DM mass and is fairly insensitive to the strength of DM self-interaction. Processes involving exclusively SOPTs, shown in the bottom row, are more common and generally require a more compact mass spectrum as well as larger values of $\lambda_2$. By comparing these two kinds of solutions, we see that processes with $\lambda_2 \lesssim 2.5$, $m_A \gtrsim 275$ GeV or $\lambda_{345} \lesssim 0.75$ necessarily involve at least one FOPT.

We can see that DM relic density increase near the Higgs resonance region ($m_H\sim m_h/2$), as expected. Furthermore, we point out that the white region for small $m_A$ is due to the collider constraints, which force $m_A \gtrsim 100$~GeV or an almost degenerate neutral scalar mass spectrum. As a final remark, let us mention that we do not find three-step PTs with three FOPT in our scan.

\subsection{Phase-transition parameters}

\begin{figure}[h!]
\begin{center}
\includegraphics[width=.8\linewidth,trim = 0cm 0cm 0cm 0cm, clip]{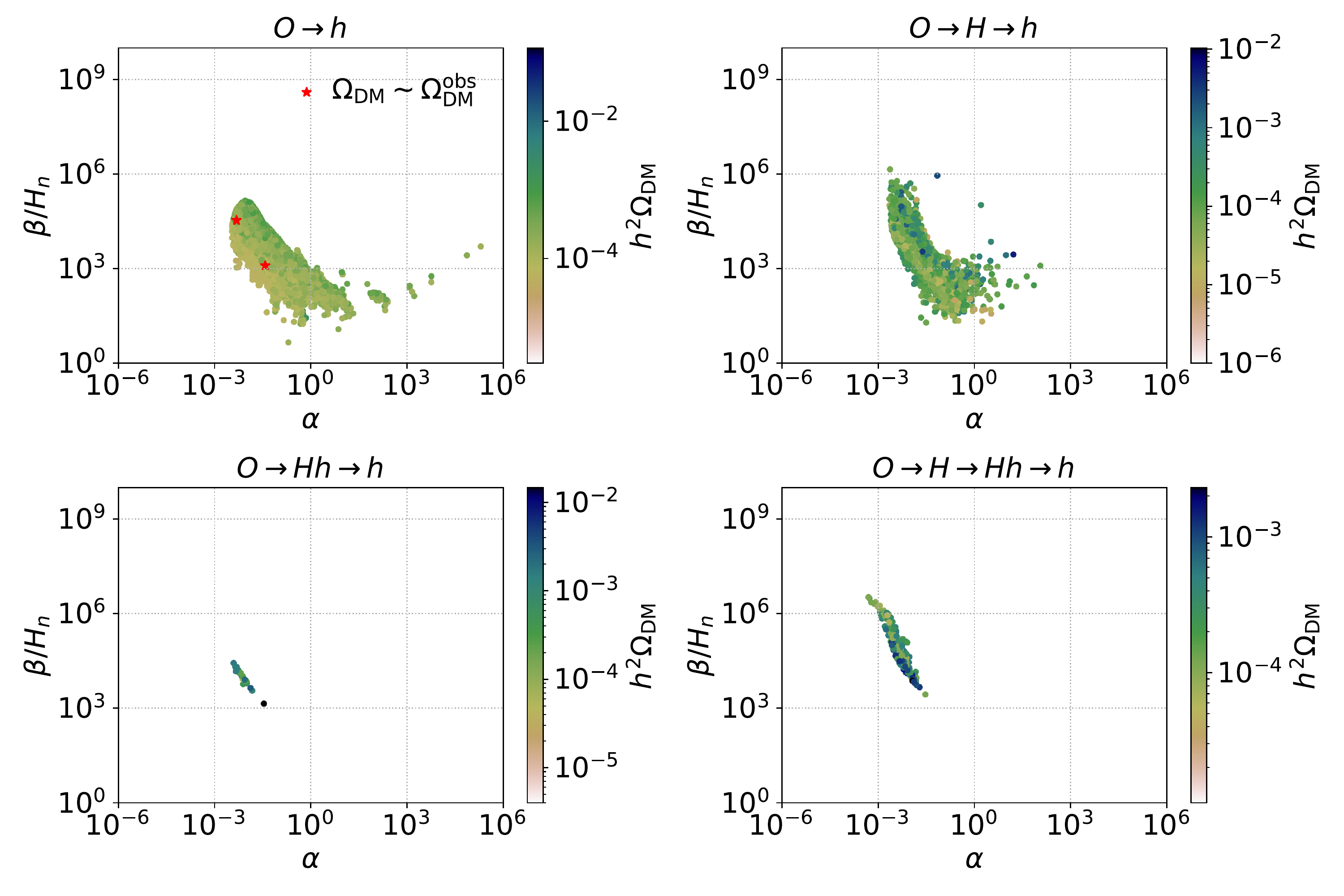}
\textit{\caption{Correlation between the phase transition parameters $\alpha$~\eqref{eq:alpha} and $\beta/H_n$~\eqref{eq:beta} in the four phase-transition patterns identified. The colour code indicates the value of the obtained DM relic abundance, which is within the Planck $3\sigma$ range where marked with red stars. Points leading to DM overabundance are not shown.}\label{fig:betaH_alpha}}
\end{center}
\end{figure}

To conclude, we study correlations between the parameters that characterise the obtained PTs. Fig.~\ref{fig:betaH_alpha} shows that all first-order phase transitions establish the usual relation between $\beta/H_n$ and $\alpha$, since the former is proportional to $1/\sqrt{\Delta V}$, while the latter scales as $\Delta V$ \cite{Grojean:2006bp}. In fact, the shorter the FOPT is (large $\beta/H_n$), the weaker it is (small $\alpha$) and vice versa. 
Note that a large value of $\beta/H_n$ signals that the PT dynamics proceeds much faster than the expansion of the Universe and, therefore, that the latter can be safely neglected. We also observe that the strongest (and thus the longest) FOPTs take place in the one-step $O\rightarrow h$ pattern.

\begin{figure}[h!]
\begin{center}
\includegraphics[width=.8\linewidth,trim = 0cm 0cm 0cm 0cm, clip]{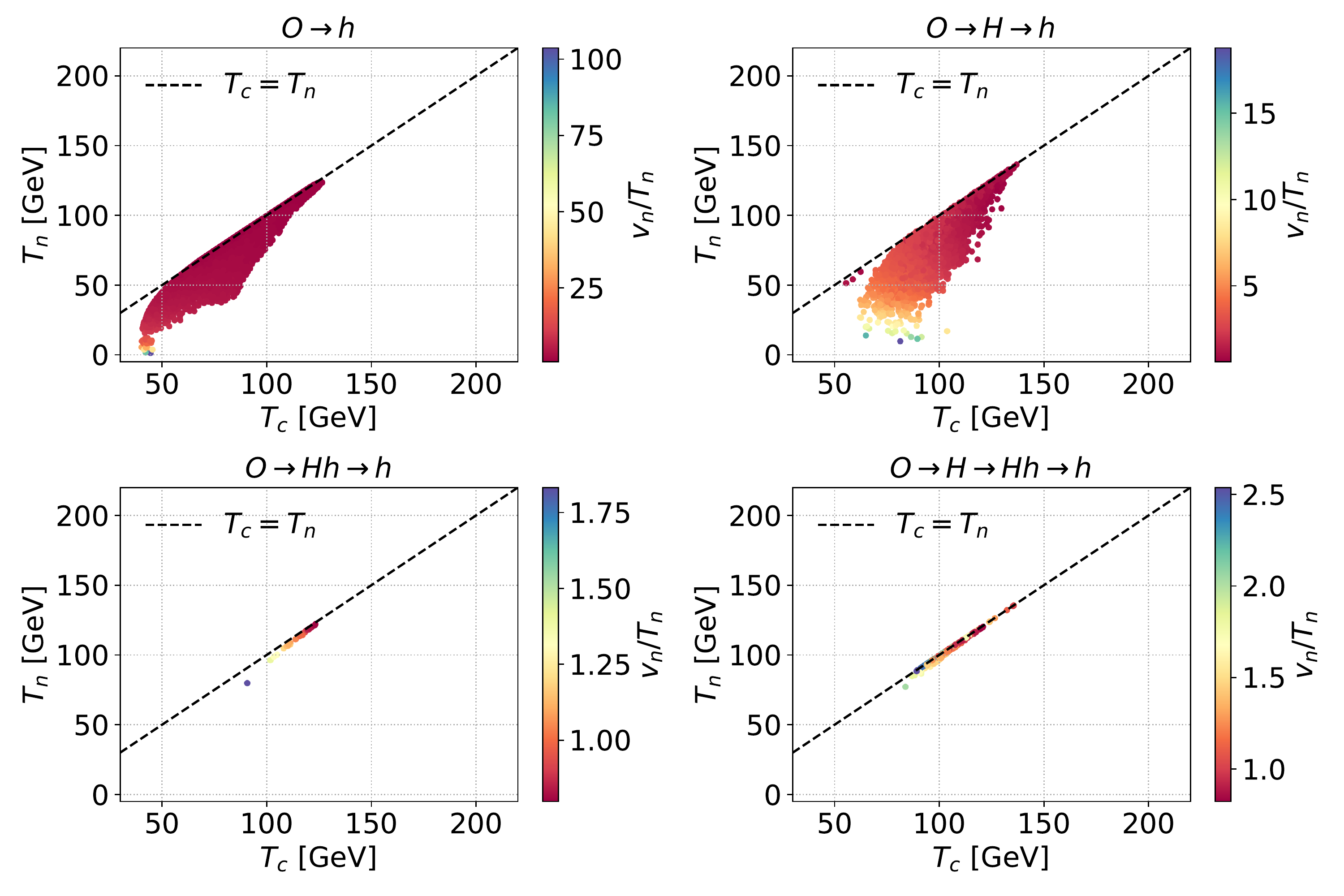}
\textit{\caption{Correlation between the nucleation temperature, $T_n$, and the critical temperature, $T_c$, for the four phase-transition patterns identified. The colour code indicates the value of the ratio $v_n/T_n$, with $v_n\equiv \sqrt{\langle H\rangle_n^2+\langle h\rangle_n^2}$ being the VEV of the true vacuum at the nucleation temperature. The dashed line indicates where the critical and nucleation temperature are equal. Points leading to DM overabundance are not shown.}\label{fig:Tc_Tn}}
\end{center}
\end{figure}

Finally, we look for correlation in the $T_c-T_n$ space shown in Fig.~\ref{fig:Tc_Tn}. Most of the points yield low values of $v_n/T_n$, and larger values of $v_n/T_n$ are only found for low $T_c$ and $T_n$, with $v_n\equiv \sqrt{\langle H\rangle_n^2+\langle h\rangle_n^2}$ being the VEV of the true vacuum at the nucleation temperature. Indeed, the latter generally increases when $T_n$ decreases, and thus $v_n/T_n$ increases as well. This behaviour can be intuitively explained as follows: for large $T_n$, $v_n$ can be small and subsequently evolve smoothly toward its final value as the temperature decreases; by contrast, small $T_n$ forces larger $v_n$ as the system has less time left to evolve smoothly toward its tree-level value. The range of values for $v_n/T_n$  is much smaller for multi-step phase transitions because the minimal value found for $T_n$ is 
larger than in the one-step case.  Furthermore, in each panel, most of the points fall close to the dashed line signalling where $T_c=T_n$. For these points, the transition takes place close to the moment where the false and true vacua become degenerate (at $T_c$), implying that a weak FOPT (small $\alpha$) with negligible supercooling drives the process. The phase transitions become stronger ($T_c-T_n$ increases) as the temperature decreases.

\section{Impact of direct detection experiments and future collider searches}
\label{sec:dd_and_monojet}

\begin{figure}[h!]
\begin{center}
\includegraphics[width=0.495\textwidth,trim = 0cm 0cm 0cm 0cm, clip]{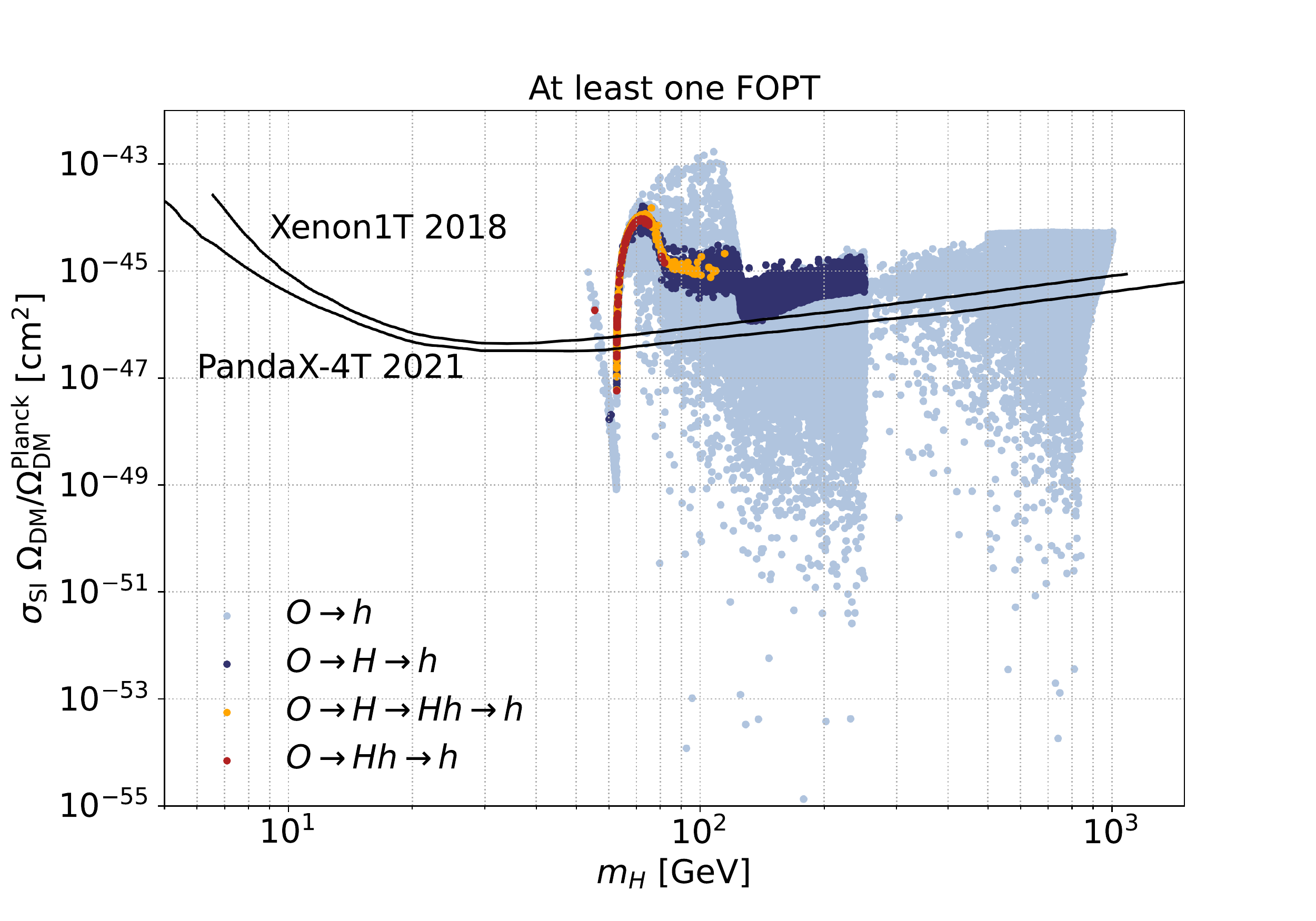}
\includegraphics[width=0.495\textwidth,trim = 0cm 0cm 0cm 0cm, clip]{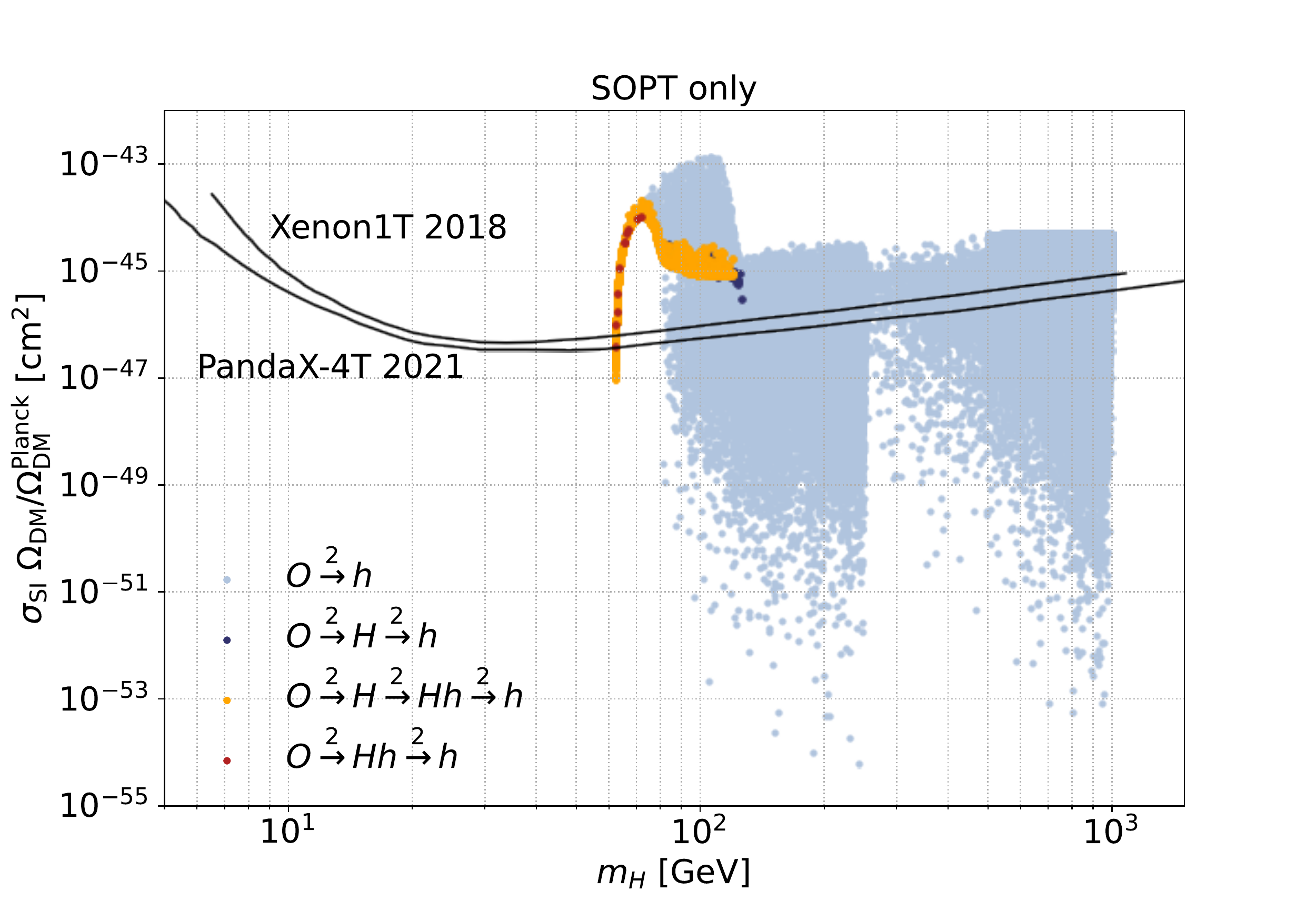}
\textit{\caption{Spin-independent direct-detection cross section as function of the DM mass for transition patterns involving at least one FOPT (left panel) or exclusively SOPTs (right panel). The colour code highlights the pattern type: one-step PT $O\rightarrow h$ (light blue),  two-step PT $O\rightarrow H\rightarrow h$ (dark blue) and $O\rightarrow Hh \rightarrow h$ (red), as well as a three-step PT $O\rightarrow H\rightarrow Hh\rightarrow h$ (orange). The experimental bounds are taken from Refs.~\cite{XENON:2018voc,PandaX-4T:2021bab}. Points leading to DM overabundance are not shown. }\label{fig:DD_cross_section}}
\end{center}
\end{figure}

Before proceeding with the analysis of the resulting GW signal, we consider a further bound given by the direct detection experiments, which probe the spin-independent cross section of DM on nuclei. To this purpose, we show in Fig.~\ref{fig:DD_cross_section} the obtained spin-independent scattering cross section $\sigma_{\rm SI}$ as function of the DM mass, highlighting the different transition pattern identified. The analysis is presented separately for processes involving at least one FOPT (left panel) or exclusively SOPTs (right panel). Because we allow for DM under-abundances, the plots have been obtained by re-scaling the cross section with the fraction $\Omega_{\rm DM}/\Omega_{\rm DM}^{\rm Planck}$, where $\Omega_{\rm DM}$ is the DM abundance produced by the IDM and $\Omega_{\rm DM}^{\rm Planck}$ the value given by the latest Planck measurement~\cite{Planck:2018vyg}. The indicated experimental bounds use the 2018 release of the XENON1T data~\cite{XENON:2018voc} and the 2021 PandaX-4T result~\cite{PandaX-4T:2021bab}. 

As we can see, most of the multi-step PTs fall above of the considered exclusion bounds in both the cases. These processes may still occur near the Higgs resonance region ($m_H\simeq m_h/2$), where resonance effects allow for the lower values of the $\lambda_{345}$ coupling required by these solutions. Another region of interest is for $m_H\in [120, 160]$ GeV, precluded for solutions involving only SOPTs and resulting in a signal borderline with the current exclusions for processes involving at least one FOPT. Contrary to the Higgs resonance region, these solutions select only multi-step PTs following the pattern $O\rightarrow H\rightarrow h$ and yield underabundant DM.


\begin{figure}[h!]
\begin{center}
\includegraphics[width=.99\linewidth]{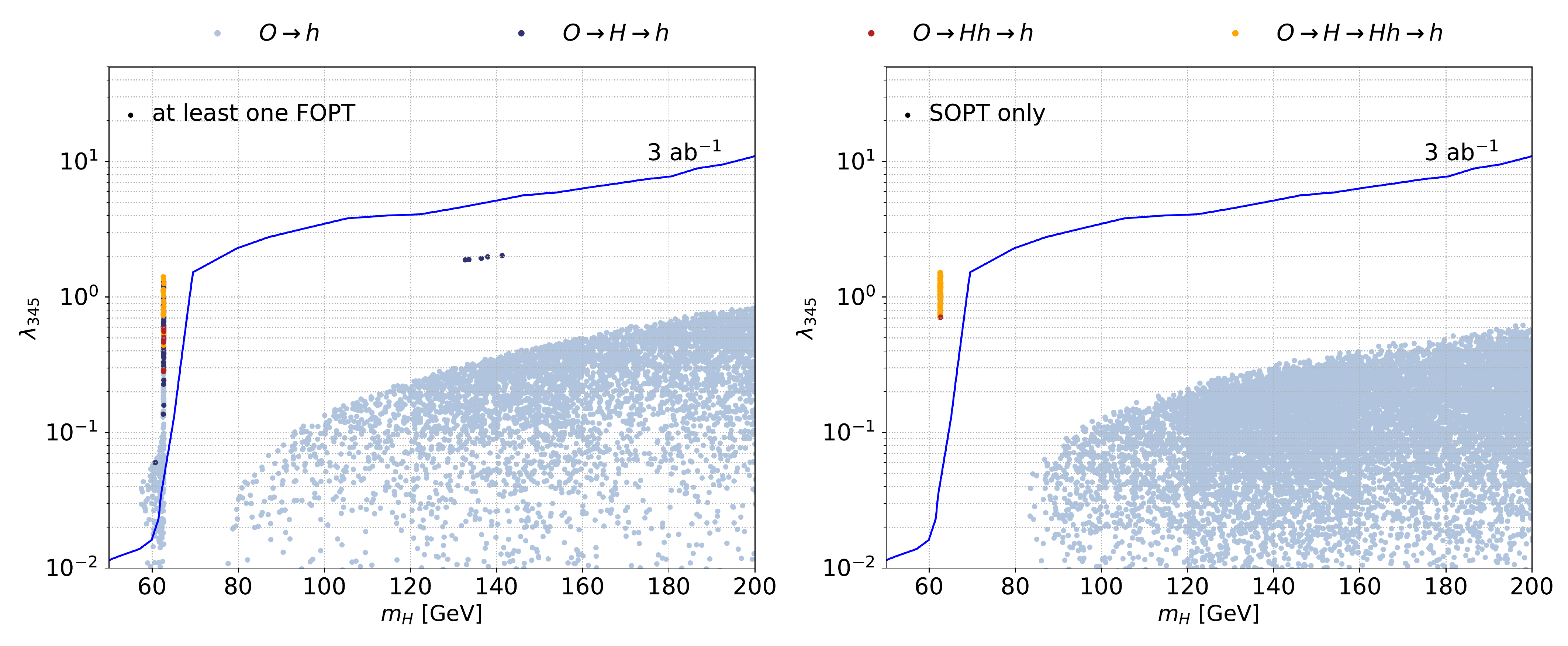}
\textit{\caption{Power of future monojet searches~\cite{Belyaev:2016lok, Belyaev:2018ext} to test IDM transition patterns involving at least one FOPT (left panel) or exclusively SOPTs (right panel). The colour code highlight the pattern type: one-step PT $O\rightarrow h$ (light blue),  two-step PT $O\rightarrow H\rightarrow h$ (dark blue) and $O\rightarrow Hh \rightarrow h$ (red), as well as a three-step PT $O\rightarrow H\rightarrow Hh\rightarrow h$ (orange). Points leading to DM overabundance and excluded by Xenon1T are not shown.}\label{fig:monojet}}
\end{center}
\end{figure}

The bounds of monojet searches, that target the production of DM at colliders, are currently less stringent than the exclusions due to direct detection experiment and Higgs measurements~\cite{Belyaev:2016lok, Belyaev:2018ext}. Nevertheless, the future LHC data has the potential to competitively probe the low end of the allowed DM mass range with these searches. In our analysis, we use the available projections of the monojet reach to highlight the phase transition patterns and the GW signal supported by the parameter space accessible to the high luminosity (HL) LHC run. 

The results obtained are shown in Fig.~\ref{fig:monojet}, where again we distinguish between patterns with at least one FOPT (left panel) and those proceeding purely through SOPTs (right panel). As we can see, the HL-LHC monojet searches have the power to completely probe IDM multi-step PT, barring a small region around $m_H\simeq 130$ GeV accessible only to patterns with at least one FOPT. This region, however, will likely be probed in direct detection experiments. Disregarding these borderline solutions, we therefore conclude that future monojet searches are crucial for establishing the viability of multi-step phase transitions within the IDM. In case of a positive detection, these experiments would also deliver a clear prediction concerning the presence of a GW counterpart to the collider signal.    

\section{Gravitational wave signal}
\label{sec:gw}

\subsection{Contributions to the stochastic GW background}

The stochastic GW background is generated by the dynamics of the nucleated vacuum bubbles, which shape the GW power spectrum $h^2\Omega_{\rm GW}$ through three processes: bubble collisions, the subsequent propagation of sound waves and through magnetohydrodynamic (MHD) turbulence. The full signal is the sum of these three contributions: $h^2\Omega_{\rm GW} \simeq h^2\Omega_{\rm col} + h^2\Omega_{\rm sw} + h^2\Omega_{\rm turb}$~\cite{Caprini:2015zlo}.

The emission of GW in bubble collisions can be quantified as a function of the frequency $f$ in~\cite{Caprini:2015zlo}
\begin{equation}
h^2\Omega_{\rm col}(f) = 1.67 \times 10^{-5}   \, \left( \frac{H_*}{\beta} \right)^2 \left( \frac{\kappa_\phi \alpha}{1+\alpha} \right)^2  
 \left( \frac{100}{g_*} \right)^{\frac{1}{3}} \left(\frac{0.11\,v_w^3}{0.42+v_w^2}\right) \, S_{\rm env}(f) \, ,
\label{eq:Omenv}
\end{equation}
where
\begin{equation}
\label{Senv}
S_{\rm env}( f ) = \frac{3.8 \,\,(f/f_{\rm env})^{2.8}}{1 + 2.8 \, (f/f_{\rm env})^{3.8}}
\end{equation}
parametrises the spectral shape of the GW radiation~\cite{Huber:2008hg}, $\kappa_\phi$ denotes the fraction of latent heat transformed into the kinetic energy of the scalar field, $v_w$ is the wall velocity, and where the frequency at the peak is defined as~\cite{Caprini:2015zlo}
\begin{equation}
f_{\rm env} = 16.5\times 10^{-6}\text{~Hz}\left(\frac{0.62}{1.8-0.1v_w+v_w^2}\right)\left(\frac{\beta}{H_*}\right)\left(\frac{T_*}{100\text{~GeV}}\right)\left(\frac{g_*}{100}\right)^{1/6}.
\label{eq:fenv}
\end{equation}

The sound wave contribution is given by~\cite{Hindmarsh:2017gnf, Caprini:2019egz, Schmitz:2020rag} 
\begin{equation}
\begin{aligned}
h^2\Omega_{\text{sw}}(f) = h^2 ~7.24\times 10^{-2} F_\text{gw,0}\left( \frac{\kappa_{\rm sw} \alpha}{1+\alpha} \right)^2  \left(\frac{H_*}{\beta}\right) \frac{\text{max\{$c_s, v_w$\} }}{c_s} \Upsilon S_{\rm sw}(f) \, ,
\label{eq:OmGsound} 
\end{aligned}
\end{equation}
where
\begin{equation}
    S_{\rm sw}( f ) = (f/f_{\rm sw})^{3}\,\left(\frac{7}{4 + 3\,(f/f_{\rm sw})^{2}}  \right)^{7/2},
\end{equation}
$c_s = 1/\sqrt{3}$ is the speed of sound in the plasma~\cite{Espinosa:2010hh}, the efficiency $\kappa_{\rm sw}$ denotes the fraction of latent heat that is transformed into bulk motion of the fluid and the parameter $F_\text{gw,0}$ is defined as
\begin{equation}
F_\text{gw,0} = \Omega_{\gamma,0}\left(\frac{g_{s0}}{g_{s*}}\right)^{4/3}\frac{g_*}{g_0},
\end{equation}
with $\Omega_{\gamma}$ photon density parameter and $g_{s}$ the effective number of entropic degrees of freedom. The subscript 0 signals that the parameters are evaluated at present values. Finally, the suppression factor $\Upsilon$ accounts for the finite lifetime of the sound waves~\cite{Caldwell:2022qsj,Guo:2020grp,Guo:2021qcq}. Again $f_{\rm sw}$ is the peak frequency, which is defined as
\begin{equation}
f_{\rm sw} = 8.88\times 10^{-6}\text{~Hz}\left(\frac{\beta}{H_*}\right) \frac{1}{\text{max\{$c_s, v_w$\} }}\left(\frac{T_*}{100\text{~GeV}}\right)\left(\frac{g_*}{100}\right)^{1/6}.
\label{eq:fsw}
\end{equation}

For the turbulence contribution we use  
\cite{Caprini:2009yp,Binetruy:2012ze} 
\begin{equation}
    h^2\Omega_{\rm turb}(f) = 
3.35 \times 10^{-4} \, \left( \frac{H_*}{\beta} \right)
\left(\frac{\kappa_{\rm turb}\,\alpha}{1+\alpha}\right)^{\frac{3}{2}}\,
 \left( \frac{100}{g_*}\right)^{1/3}\, v_w \, S_{\rm turb} (f) \, ,
 \label{eq:turb1}
\end{equation}

where the spectral shape is given by~\cite{Caprini:2009yp,Binetruy:2012ze} 
\begin{equation}
S_{\rm turb} (f) = \frac{(f/f_{\rm turb})^3}
{\left[ 1 + (f/f_{\rm turb}) \right]^{\frac{11}{3}} 
\left(1 + 8 \pi f/h_* \right)} \,
\label{Sturb},
\end{equation}
$\kappa_{\rm turb}$ denotes the fraction of latent heat that is transformed 
into MHD turbulence and $h_*$ is given by~\cite{Caprini:2015zlo}
\begin{equation}
    h_* = 16.5\times 10^{-6}\text{~Hz}\left(\frac{T_*}{100\text{~GeV}}\right)\left(\frac{g_*}{100}\right)^{1/6}.
\end{equation}
Notice that the parameter $f_{\rm turb}$, determined by~\cite{Caprini:2015zlo}
\begin{equation}
f_{\rm turb} = 2.7\times 10^{-5}\text{~Hz}\left(\frac{\beta}{H_*}\right) \frac{1}{v_w}\left(\frac{T_*}{100\text{~GeV}}\right)\left(\frac{g_*}{100}\right)^{1/6}
\label{eq:fturb}
\end{equation}
does not represent the peak frequency of the corresponding contribution.\footnote{This would be the case if $h^2\Omega_{\rm turb}$ and $S_{\rm turb}$ were normalised in a different way.}

As for the wall velocity $v_w$, the precise determination of this parameter is very challenging as it requires a precise computation of the effect sourcing the friction term, as well as the solution of the Boltzmann equations that model the interaction of the scalar fields driving the PT with the plasma, see for instance~\cite{Laurent:2020gpg,Dorsch:2021ubz,Dorsch:2021nje,DeCurtis:2022hlx,Laurent:2022jrs, Lewicki:2021pgr} for recent works. As suggested by the recent analysis in Ref.~\cite{Laurent:2022jrs}, we consider fast moving walls and consequently set $v_w\simeq 1$ for simplicity.

In our computation we track the contributions $h^2\Omega_{\rm col},  h^2\Omega_{\rm sw}, h^2\Omega_{\rm turb}$ separately, finding that for the model at hand it is the sound wave component that usually dominates.

\subsection{General scan}
\label{sec:general_scan}
\begin{figure}[h!]
\begin{center}
\includegraphics[width=0.7\linewidth]{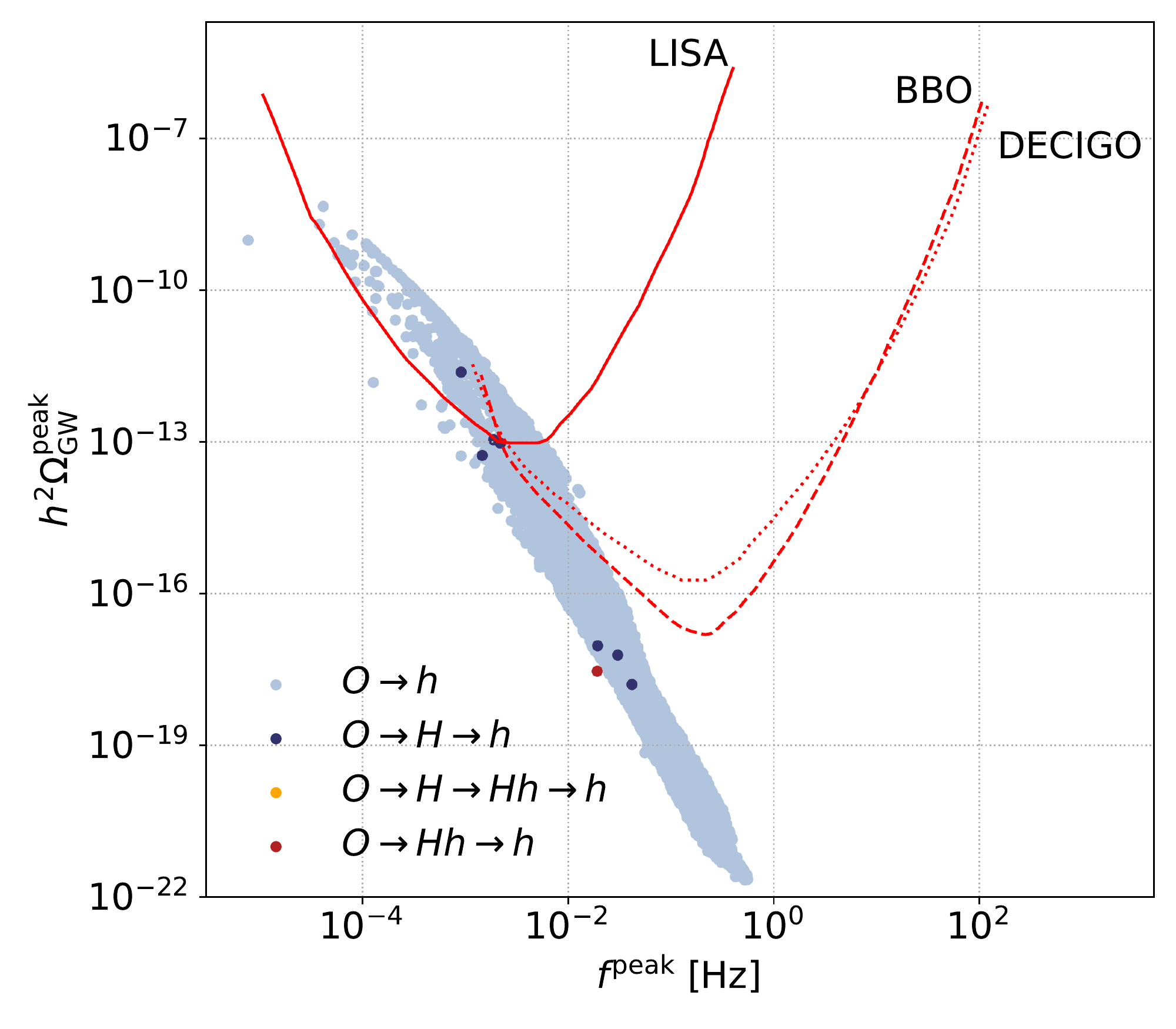}
\textit{\caption{GW signal $h^2 \Omega_{\rm GW }^{\rm peak}$ as a function of the peak frequency $f^{\rm peak}$ for the considered parameter space. The colour code indicates the PT pattern. Points leading to DM overabundance and excluded by Xenon1T are not shown. The  power-law integrated sensitivity curves of future gravitational wave detectors are also shown.}  \label{fig:GW_all_contributions}}
\end{center}
\end{figure}

The GW signals supported by the IDM parameter space are shown in Fig.~\ref{fig:GW_all_contributions}, where we depict the value of the 
the peak of the power spectrum $h^2\Omega_{\rm GW }^{\rm peak}$ and the associated frequency at this peak $f^{\rm peak}$ for each point of scan performed. 

The obtained GW signals are always dominated by the sound wave contribution. We also display the power-law integrated  sensitivity curves of future GW detectors LISA~\cite{eLISA:2013xep,2017arXiv170200786A}, BBO~\cite{Corbin:2005ny,Crowder:2005nr} and DECIGO~\cite{Seto:2001qf,Kawamura:2020pcg}. LISA, in particular, will probe mostly one-step transitions $O\rightarrow h$ and part of the solutions using the $O\rightarrow H \rightarrow h$ pattern. 
Overall, we see that single-step PTs tend to produce stronger signals. A strong signal is correlated with a large value of $\alpha$, which, as shown in Fig.~\ref{fig:betaH_alpha}, implies a smaller value for $\beta/H_n$. Moreover, the frequency at the GW power spectrum peak is inversely proportional to the characteristic time scale of the PT, $\beta^{-1}$.\footnote{As shown in Eqs.~(\ref{eq:Omenv}), (\ref{eq:OmGsound}) and (\ref{eq:turb1}), each contribution in  $h^2\Omega_{\rm GW}$ is multiplied by a factor of $H_n/\beta$ (this factor is squared for bubble collisions), while in the relation for $f^{\rm peak}$, Eq.~(\ref{eq:fenv}), (\ref{eq:fsw}) and (\ref{eq:fturb}) show that a factor of $\beta/H_n$ appears in each term.} This explains the trend in Fig.~\ref{fig:GW_all_contributions}. 

While few points yielding multiple FOPTs passed our selection criteria, we note that the corresponding GW signals are generally strongly dominated by the dynamics of one of the steps. For these solutions, FOPTs initiated in the $O$ phase are weaker than those starting from an $H$ or $Hh$ phase mainly because of the difference in the released vacuum energy $\Delta V_{\rm eff}$, which can be several orders of magnitudes larger in the latter cases. We also observe a difference in $T_n$, but this is usually not large enough for $\rho_{\rm rad}\sim T^4$ to play a significant role in the discrepancies. 

\begin{figure}[h!]
\begin{center}
\includegraphics[width=0.7\linewidth]{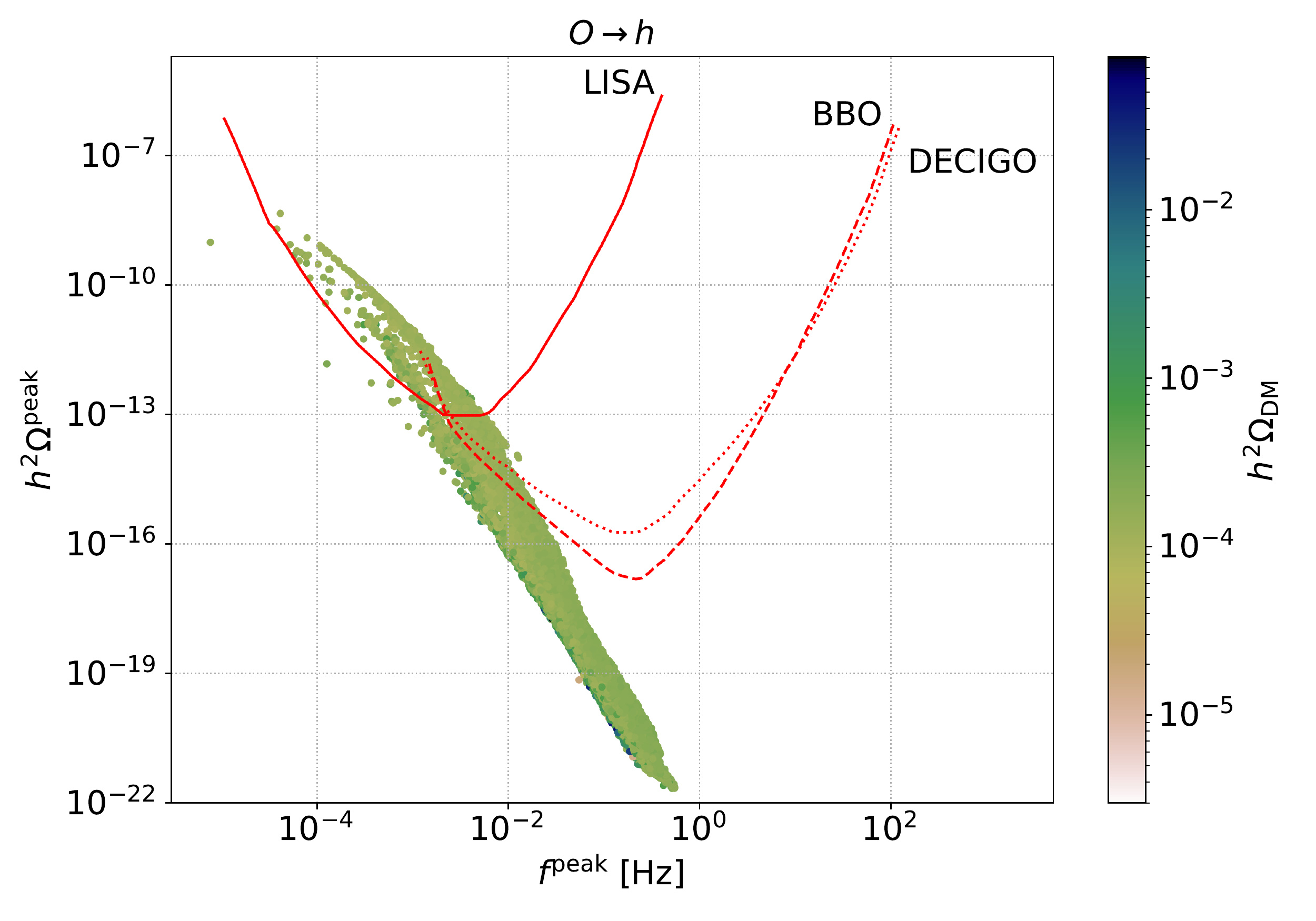}
\textit{\caption{GW signal $h^2 \Omega_{\rm GW }^{\rm peak}$ as function of the peak frequency $f^{\rm peak}$ for one-step FOPT $O\rightarrow h$. The colour code indicates the obtained DM relic density. The power-law integrated sensitivity curves of future gravitational wave detectors are also shown.}
\label{fig:GW_f_relic_density_o_h}}
\end{center}
\end{figure}

The analysis of the GW signal is repeated in Fig.~\ref{fig:GW_f_relic_density_o_h}, where we seek correlations with the DM relic abundance produced by the IDM focusing on one-step PT, which contribute to the bulk of the signal. As we can see, the IDM parameter space resulting in a DM relic abundance nearly compatible with the Planck measurements, spanned mostly by solutions with $m_H \simeq m_h/2$ and $0<\lambda_{345} \leq 0.05$, remains well below the forecast sensitivities of the analysed experiments. For these ranges of $m_H$ and $\lambda_{345}$, we only find one-step FOPT. 

\begin{figure}[h!]
\begin{center}
\includegraphics[width=.99\linewidth]{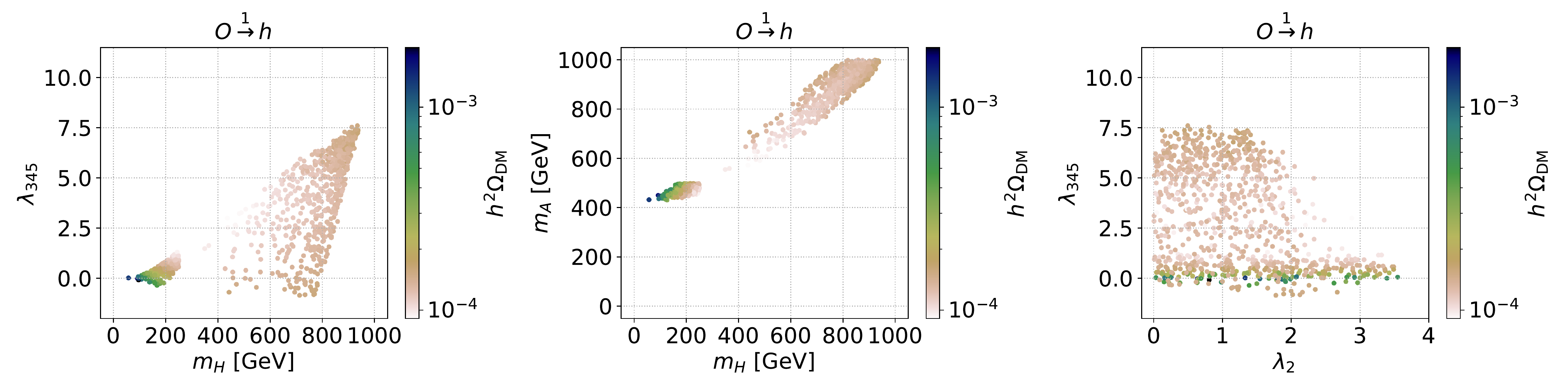}
\textit{\caption{IDM parameter space yielding a gravitational wave signal detectable at LISA~\cite{eLISA:2013xep,2017arXiv170200786A}. The colour code indicates the produced DM abundance. Points leading to DM overabundance and excluded by Xenon1T are not shown.}
\label{fig:lisaPS}}
\end{center}
\end{figure}

In order to assess the power of the LISA experiment to probe the IDM parameter space, in Fig.~\ref{fig:lisaPS} we have retained only one-step transitions sourcing a detectable signal.  We notice the presence of two kinds of solutions: the first one selects lighter scalar masses, $m_H \lesssim 250$ GeV and $m_A\lesssim 500$ GeV, and requires limited value of the $\lambda_{345}\lesssim 2$ coupling regardless of the magnitude of $\lambda_2$. A second class, instead, is given by larger values of the masses and of the $\lambda_{345}$ parameter, but imposes an upper bound on the DM self-interactions regulated by $\lambda_2$. As mentioned before, both of these solutions lead only to an underabundant DM relic density. The successful detection by LISA of a signal compatible with the IDM would, therefore, require the presence of other DM candidates besides the inert one.  

\subsection{Benchmark points}

For closer study, we analyse a number of benchmark points obtained from the configuration $m_A=400$ GeV, $m_{H^+}=235$ GeV, $\lambda_2 =3$ and $\lambda_{345}=0.6$ by varying $m_H$. As shown in Tab.~\ref{tab:BM}, the considered range of $m_H$ scans across different PT patterns: lower masses give three-step PTs, while masses above about 70 GeV yield a single-step PT. The fraction of produced DM relic abundance progressively diminishes with increasing values of $m_H$, stabilising to the permille level in the high end of the considered range.  

\begin{table}[h!]
\begin{center} 
\begin{tabular}{|c|c|c|c|c|c|} 
 \hline
 $m_H$ (GeV) & Phase transition & $T_c$ (GeV) &  $\alpha$ & $\beta/H_n$ & $\Omega_{\mathrm{DM}}/\Omega_{\mathrm{DM}}^{\rm Planck}$\\ 
 \hline
 & $\mathbf{O \overset{2}{\rightarrow} H} \overset{1}{\rightarrow} Hh \overset{2}{\rightarrow}h$ & 158.62&--& --& \\
10 & $O \overset{2}{\rightarrow} \mathbf{H \overset{1}{\rightarrow} Hh} \overset{2}{\rightarrow}h$ & 112.59 & $2.82\times 10^{-3}$& $8.00\times 10^4$ & 0.106\\
 & $O \overset{2}{\rightarrow} H \overset{1}{\rightarrow} \mathbf{Hh \overset{2}{\rightarrow}h}$ & 17.09&--&--& \\
\vdots & \vdots & \vdots& \vdots&\vdots&\vdots\\
 & $\mathbf{O \overset{2}{\rightarrow} H} \overset{1}{\rightarrow} Hh \overset{2}{\rightarrow}h$ & 148.93&--&--&\\
40 & $O \overset{2}{\rightarrow} \mathbf{H \overset{1}{\rightarrow} Hh} \overset{2}{\rightarrow}h$ & 113.02 &$3.60\times 10^{-3}$&$5.01\times 10^4$& 0.038\\
 & $O \overset{2}{\rightarrow}H \overset{1}{\rightarrow}  \mathbf{Hh \overset{2}{\rightarrow}h}$ & 73.20&--&--&\\
  & $\mathbf{O \overset{1}{\rightarrow} H} \overset{1}{\rightarrow} Hh \overset{2}{\rightarrow}h$ & 142.70& $1.06\times 10^{-6}$ & $2.33 \times 10^8$&\\
  50 & $O \overset{1}{\rightarrow} \mathbf{H \overset{1}{\rightarrow} Hh} \overset{2}{\rightarrow}h$ & 113.68 & $4.03\times 10^{-3}$& $4.00\times 10^4$& 0.014\\
  & $O \overset{1}{\rightarrow} H \overset{1}{\rightarrow} \mathbf{Hh \overset{2}{\rightarrow}h}$ & 85.73&--&--&\\
 \vdots & \vdots & \vdots&\vdots&\vdots&\vdots\\
  &$\mathbf{O \overset{1}{\rightarrow} H} \overset{1}{\rightarrow} Hh \overset{2}{\rightarrow}h$ & 121.13&$1.26\times 10^{-5}$ & $6.37\times 10^6$&\\
  70 &$O \overset{1}{\rightarrow} \mathbf{H \overset{1}{\rightarrow} Hh} \overset{2}{\rightarrow}h$ & 115.45 &$5.43\times 10^{-3}$&$2.16\times 10^4$& 0.004\\
  &$O \overset{1}{\rightarrow} H \overset{1}{\rightarrow} \mathbf{Hh \overset{2}{\rightarrow}h}$ & 107.36&--&--&\\
 80 & $\mathbf{O \overset{1}{\rightarrow} h}$ & 116.32 &$6.65\times 10^{-3}$&$1.13\times 10^4$& 0.001\\
\vdots & \vdots& \vdots & \vdots&\vdots&\vdots\\
210 & $\mathbf{O \overset{1}{\rightarrow} h}$ & 138.47 &$3.39 \times 10^{-4}$&$2.32\times 10^3$& 0.002\\
220 & $\mathbf{O \overset{2}{\rightarrow} h}$ & 139.84&--&--& 0.002\\
230 & $\mathbf{O \overset{2}{\rightarrow} h}$ & 141.36&--&--& 0.002\\
 \hline
\end{tabular}
\end{center}
\textit{
\caption{Benchmark points: $m_A=400~\mathrm{GeV}$, $m_{H^+}=235~\mathrm{GeV}$, $\lambda_2 =3$, $\lambda_{345}=0.6$ and varying $m_H$. The quantity $\Omega_{\mathrm{DM}}/\Omega_{\mathrm{DM}}^{\rm Planck}$ is the fraction that the IDM contributes to the observed DM relic density. For each line, the PT step in bold is the step detailed in this line.  \label{tab:BM}}}
\end{table}

\begin{figure}[h!]
\centering
\includegraphics[width=.495\linewidth]{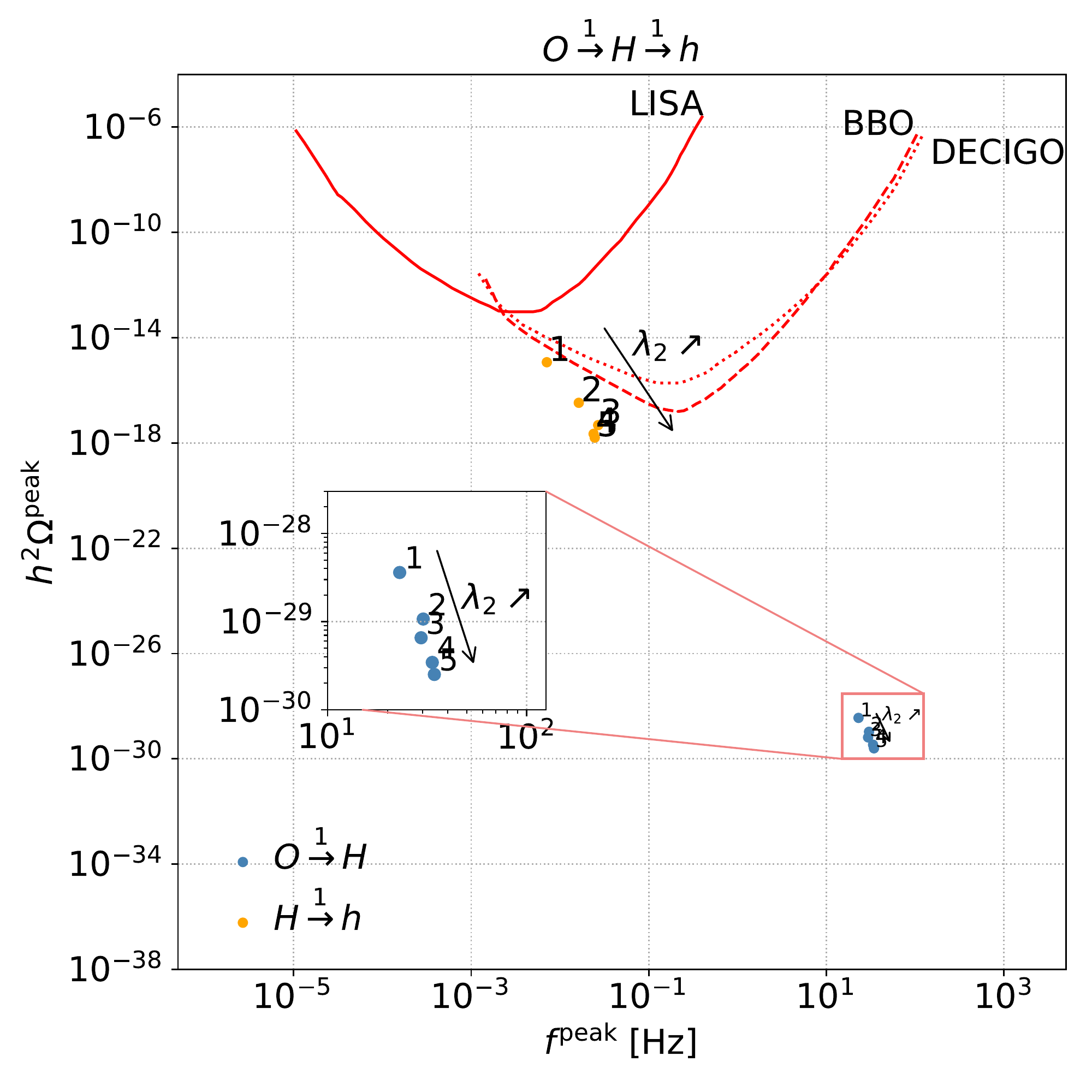}
\includegraphics[width=.495\linewidth]{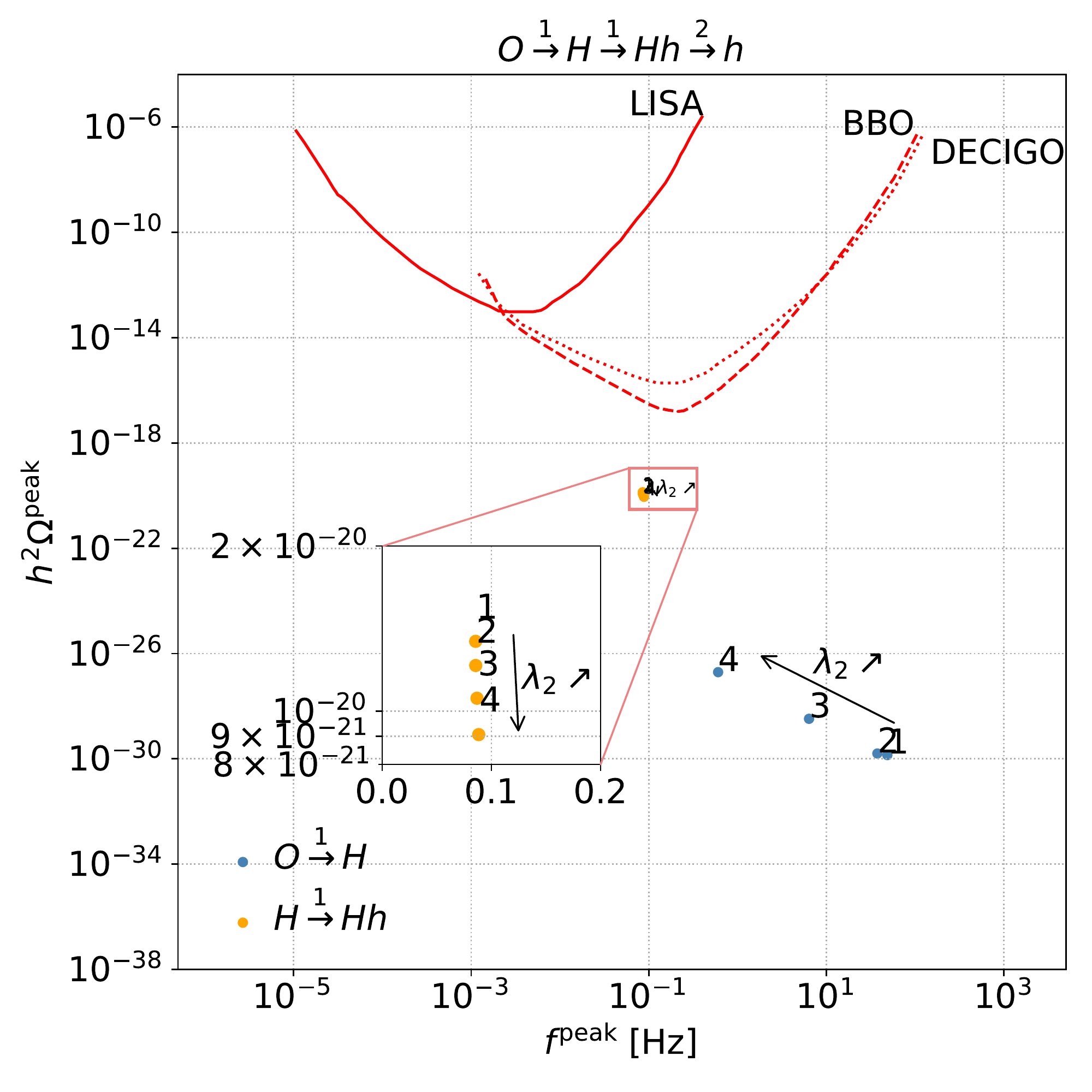}
\textit{\caption{Left panel: Gravitational wave signal $h^2 \Omega^{\rm peak}$ plotted as function of the peak frequency $f^{\rm peak}$ for a benchmark 2-step PT with $m_H=70~\mathrm{GeV}$, $m_A=400~\mathrm{GeV}$, $m_{H^+}=235~\mathrm{GeV}$, $\lambda_{345}=0.6$ and $\lambda_2\in [0.7, 1.1]$. Right panel: The same signal for a benchmark 3-step PT with $m_H=68~\mathrm{GeV}$, $m_A=398~\mathrm{GeV}$, $m_{H^+}=233~\mathrm{GeV}$, $\lambda_{345}=0.6$ and $\lambda_2\in [2.9, 3.2]$.}
\label{fig:GW_f_BM}}
\end{figure}

Finally, setting $m_H=70$ GeV and the remaining parameters around the values used in Tab.~\ref{tab:BM}, we checked the impact of a varying DM quartic coupling $\lambda_2$ on the strength of the GW signal. The results obtained are presented in Fig.~\ref{fig:GW_f_BM}, showing that an increase in the coupling usually causes a decrease in the signal. Moreover, as mentioned in Sec.~\ref{sec:general_scan}, we clearly see that a FOPT from the origin $O$ is generally weaker than the subsequent FOPT.

\section{Implication of the new CDF result on the $W$ boson mass}
\label{sec:wmass}

The CDF collaboration has recently released a new determination of the $W$ boson mass based on the full Tevatron data, finding~\cite{CDF:2022hxs}
\begin{equation}
    m_W^{\rm CDF}= (80.43335 \pm 0.0094) \text{ GeV}.
\end{equation}
The value is in strong tension with the corresponding result of global EW observable fits performed within the SM framework~\cite{ParticleDataGroup:2016lqr} 
\begin{equation}
    m_W^{\rm EW} = (80.357 \pm 0.006) \text{ GeV},
\end{equation}
yielding a deviation of about $7\sigma$ from the SM expectation. Importantly, the new CDF result is also departing significantly from the outcome of the analysis performed by the ATLAS collaboration on the LHC data, which finds $M_W=80.370\pm 0.0019$~GeV~\cite{ATLAS:2017rzl}. Pending new experimental inputs by the ATLAS and CMS collaboration, necessary to crosscheck the CDF claim, the anomaly seems to strongly suggest the presence of new physics contributions affecting the EW sector. In regard of this, as first pointed out in Ref.~\cite{Fan:2022dck}, the additional scalar bosons included in the IDM framework would allow to explain the new CDF result without violating the constraints imposed by the mentioned new physics searches.    

Assuming this interpretation of the $W$ mass anomaly, in the present section we reanalyse the impact that the new CDF result, if confirmed, would have on the results discussed in the previous sections. To this purpose, we repeat our scan considering the ranges of the Peskin-Takeuchi parameters shown in Tab.~\ref{tab:STUnew}, derived through a global fit of EWPOs that accounts for the new $m_W$ value~\cite{Lu:2022bgw}. 

\begin{table}[h]
    \centering
    \begin{tabular}{|c|c|c|}
    \hline
    Parameter & Result & Correlation \\
    \hline
         $S$ &  $0.06 \pm 0.10$ &   $0.90$ $(T)$, $-0.59$ $(U)$ \\
         $T$ & $0.11 \pm 0.12$ &  $-0.85$ $(U)$ \\
         $U$ & $0.14 \pm 0.09$ & \\
         \hline
    \end{tabular}
    \textit{\caption{Peskin-Takeuchi parameters~\cite{Peskin:1990zt,Peskin:1991sw} as determined by the electroweak precision observables~\cite{Lu:2022bgw} accounting for new $W$ boson mass determination by the CDF collaboration~\cite{CDF:2022hxs}.}
    \label{tab:STUnew}}
\end{table}

\begin{figure}[h!]
\begin{center}
\includegraphics[width=0.99\textwidth,trim = 0cm 0cm 0cm 0cm, clip]{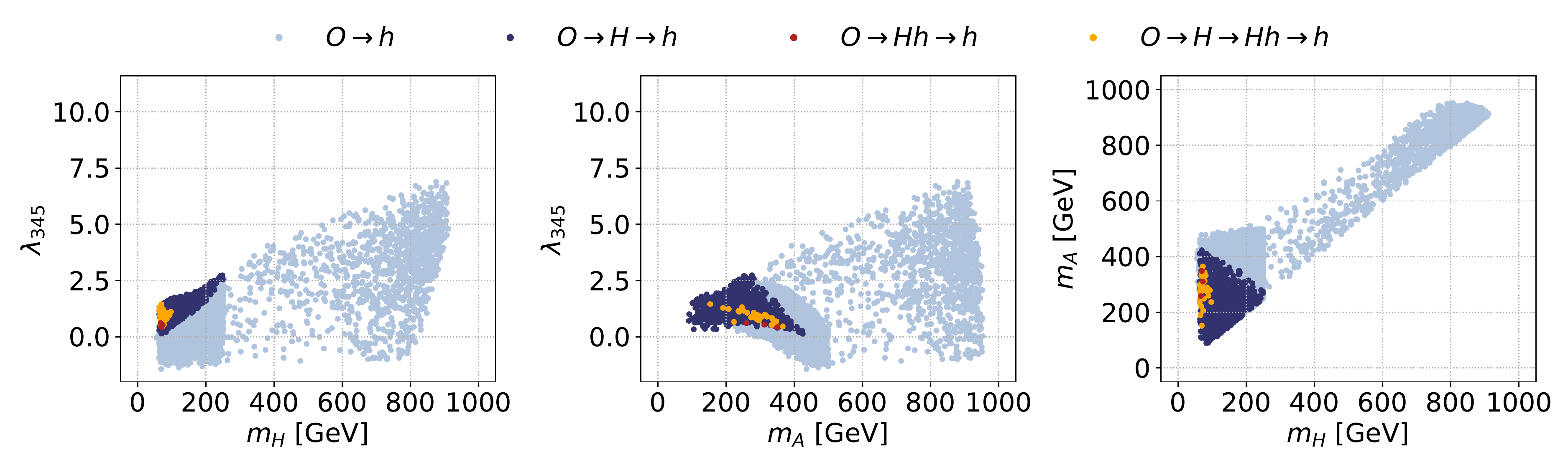}
\textit{\caption{IDM parameter space selected by the CDF result on the $W$ boson mass. All the points yield a value of $m_W$ within the $1\sigma$ range indicated by the collaboration~\cite{CDF:2022hxs}: $m_W^{\rm CDF} = (80.43335 \pm 0.0094)~\mathrm{GeV}$. The colour code denotes the PT pattern: one-step PTs $O\rightarrow h$ (light blue), two-step PTs $O\rightarrow H\rightarrow h$ (dark blue) and $O\rightarrow Hh \rightarrow h$ (red), as well as three-step PTs $O\rightarrow H\rightarrow Hh\rightarrow h$ (orange). All the transitions involve at least one FOPT. Points leading to DM overabundance are not shown. }\label{fig:PS_MW}}
\end{center}
\end{figure}

As a first step, we verify that the IDM contribution can fully explain the $W$ mass value proposed by the CDF collaboration. To this purpose, in Fig.~\ref{fig:PS_MW} we show the regions of the IDM parameter space yielding a value of $m_W$ that falls within the $1\sigma$ range of the CDF result. In order to facilitate the comparison with Fig.~\ref{fig:FOPT}, we have showed again only patterns that involve at least one first-order transition. As we can see, the new determination of $m_W$ does not preclude any of the transition patterns identified before and the distribution of the solution remains qualitatively the same. We notice the presence of an upper bound $\lambda_{345}\lesssim7$, not observed in the previous analysis, and the lack of solutions characterised by negative values of $\lambda_{345}$ and $m_H\gtrsim 800$ GeV.      

\begin{figure}[h!]
\begin{center}
\includegraphics[width=0.495\textwidth,trim = 0cm 0cm 0cm 0cm, clip]{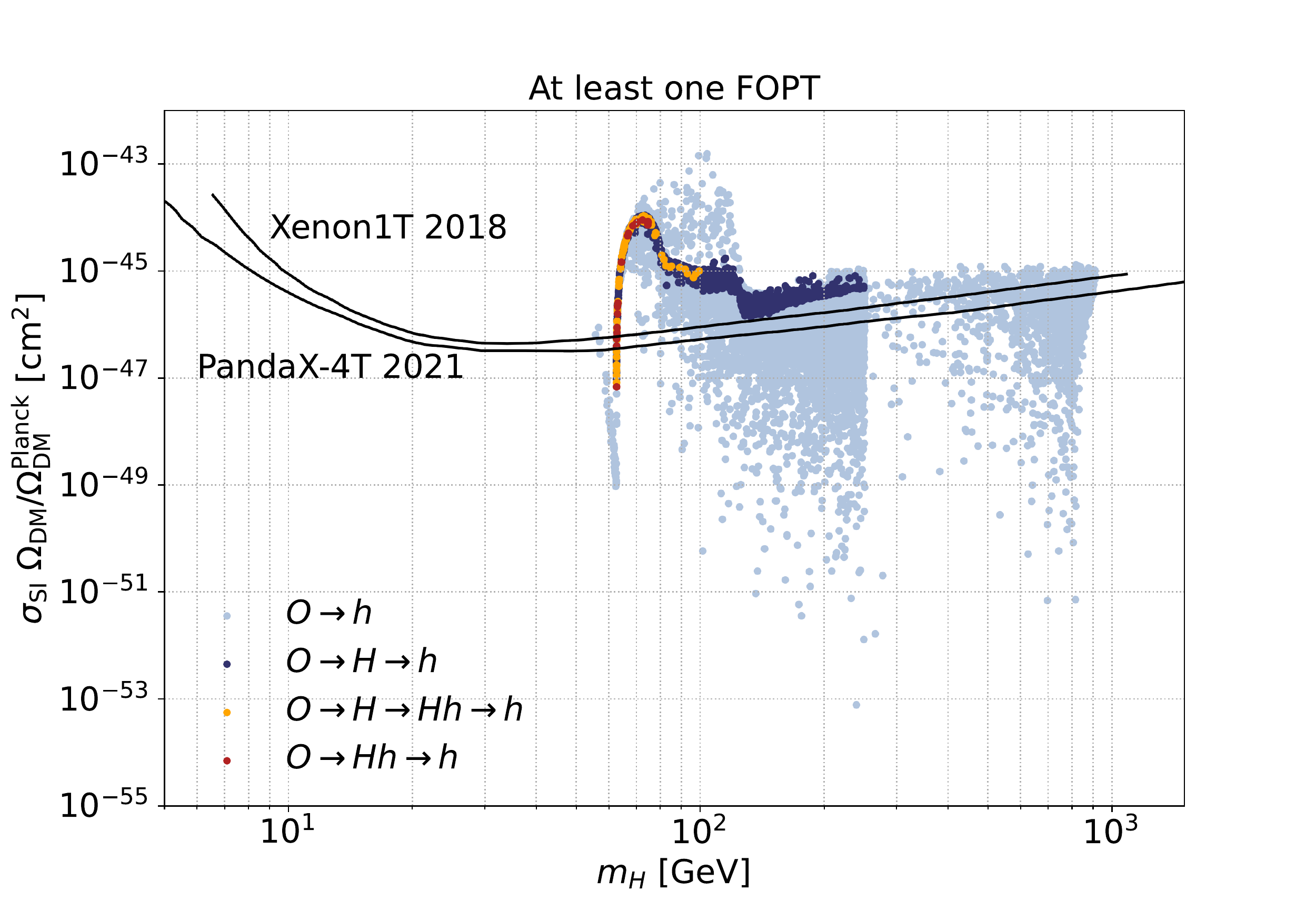}
\includegraphics[width=0.495\textwidth,trim = 0cm 0cm 0cm 0cm, clip]{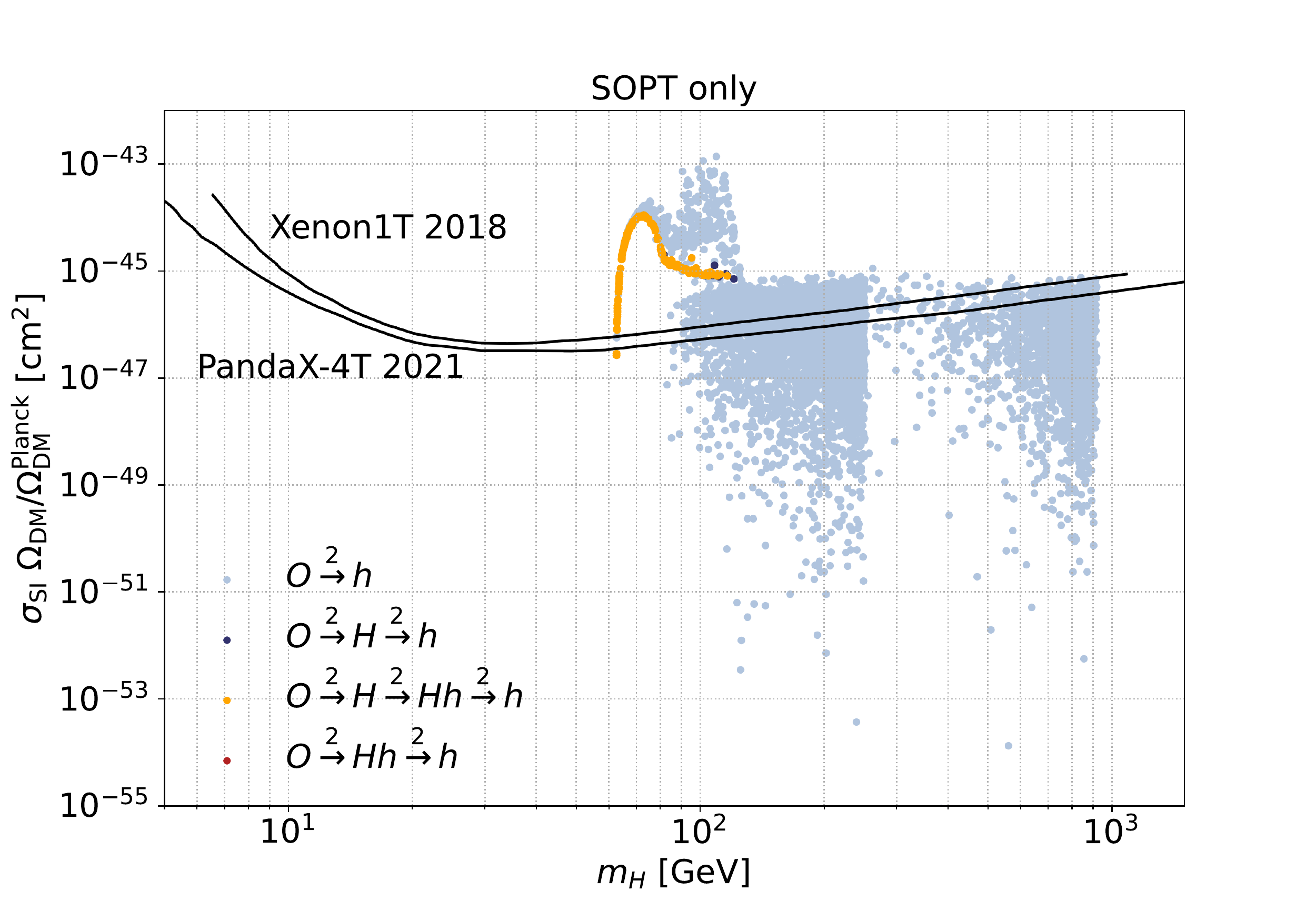}
\textit{\caption{Spin-independent direct-detection cross section as function of the DM mass for transition patterns involving at least one FOPT (left panel) or exclusively SOPTs (right panel). This plot assumes the new CDF $W$-mass result. The colour code highlights the pattern type: one-step PT $O\rightarrow h$ (light blue),  two-step PT $O\rightarrow H\rightarrow h$ (dark blue) and $O\rightarrow Hh \rightarrow h$ (red), as well as a three-step PT $O\rightarrow H\rightarrow Hh\rightarrow h$ (orange). The experimental bounds are taken from Refs.~\cite{XENON:2018voc,PandaX-4T:2021bab}. Points leading to DM overabundance are not shown.}\label{fig:DD_cross_section_newMW}}
\end{center}
\end{figure}

We continue the analysis in Fig.~\ref{fig:DD_cross_section_newMW}, where we re-analyse the results obtained for the DM spin-independent cross section on nuclei probed in direct detection experiment. As we can see, assuming the CDF determination of the $W$ boson mass strongly reduces the number of solutions that comply with the direct detection bounds for $m_H\gtrsim 100$ GeV. The effect is particularly evident for patterns with at least one FOPT (left panel), where the bulk of one-step transitions is now excluded and multi-step transitions are only allowed in the region of DM mass close to $m_h/2$. The reduction of viable solutions is also appreciable for patterns with SOPTs only (right panel), although it is less stringent. In this case, a sizeable number of one-step transitions remains well below the exclusion bounds of direct detection experiments, but multi-step transitions are essentially excluded. 

Retaining only the points allowed by the direct detection bounds and computing the implied GW signals, we find that the CDF  value of $m_W$ yields no qualitative difference with respect to the results presented in Sec.~\ref{sec:gw}. The GW spectrum is still dominated by the contribution of one-step PTs and the dependence of the signal on the peak frequency is essentially left unmodified. One minor difference pertains to the parameter space yielding a signal within the range of the LISA experiment, where we have noticed the presence of a sharper upper bound $\lambda_{345}\lesssim 6$ affecting the second class of solutions previously discussed at the end of Sec.~\ref{sec:general_scan}.


\section{Conclusions}
\label{sec:concl}

The inert doublet model is a simple extension of the Standard Model that yields a wide range of phenomenological consequences spanning from collider physics to cosmology. This framework, in particular, offers a natural dark matter candidate and predicts the appearance of additional scalar degrees of freedom at mass scales typically not much larger than the electroweak scale. Although the dark matter phenomenology and collider constraints of this model have been thoroughly studied, the cosmic phase transitions supported by its scalar potential, as well as the implied gravitational wave signals, have not been paid enough attention to. With the present paper we intended to address this issue with a comprehensive exploration of the phase structure and possible transitions supported by the inert doublet model. In our work we took into account available collider constraints,  electroweak precision observables and theoretical bounds imposed by stability of the potential and perturbativity. Furthermore, the latest results of dark matter experiments have been used to investigate the properties of the neutral scalar component of the inert doublet, assumed to provide at least a subdominant dark matter component.

Our study of the thermal evolution of the scalar potential has given a full characterization of the possible phase transition patterns supported by the inert doublet model (see Fig.~\ref{fig:phases:transs}). Although in most of the parameter space the electroweak vacuum is reached through a single phase transition, our analysis shows well-defined parameter regions where the electroweak vacuum is reached via a chain of consecutive phase transitions. 
Both two-step and three-step phase transitions with different transient phases (where only the inert doublet or both the doublets acquire a vacuum expectation value) are possible.
In Fig.~\ref{fig:SOPT_FOPT} we detail the parameter space required to have first-order phase transitions potentially detectable through gravitational wave signal. For example, a negative \emph{or} large value of the $\lambda_{345}$ quartic coupling is required to have a single-step first-order phase transition. Multi-step transitions can occur when the inert doublet components are not heavier than a few hundred GeV and couplings have moderate values, as shown in Fig.~\ref{fig:FOPT} for patterns involving at least one first-order step. After studying each of the identified transition chains separately, we show in Tab.~\ref{tab:BM}, for a benchmark point, how the dark matter mass affects the pattern and strength of the phase transitions. Fig.~\ref{fig:GW_f_BM} shows instead the dependence on dark matter self-interactions regulated by $\lambda_2$.

By cross-correlating the identified phase transition patterns with dark matter phenomenology, we find that the inert doublet model can explain the observed relic abundance only in a part of its parameter space where the electroweak vacuum is reached through single-step processes of either order. Although multi-step phase transition patterns are associated with a significant dark matter underdensity, we see that dark matter direct detection experiments are able to tightly constrain these solutions. Focusing on patterns that involve at least one first-order phase transition, Fig.~\ref{fig:DD_cross_section} shows that the direct detection bounds allow for multi-step phase transitions almost exclusively for dark matter masses close to half the Higgs boson mass. Importantly, as shown in Fig.~\ref{fig:monojet}, these solutions can be fully tested with the upcoming  high-luminosity LHC monojet searches.    

After applying the results of direct detection searches as a further constraint, we have investigated the gravitational wave spectra produced by different phase transition patterns. The results, gathered in Fig.~\ref{fig:GW_all_contributions}, show that one-step processes dominate the signal. Future gravitational wave experiments will probe a part of these solutions yielding a significant dark matter underdensity, implying that the detection of a compatible signal would require another dark matter component. Focusing on the reach of the LISA experiment, we have shown in Fig.~\ref{fig:lisaPS} the part of the inert doublet model parameter space which LISA is able to constrain. Whereas few points with multiple first-order phase transitions fall above the sensitivity curves of the considered experiments, we find that the generated gravitational signal is always strongly dominated by the transitions initiated during the transient phase at intermediate temperature. Therefore, it is highly unlikely that such transitions will induce a gravitational wave signal with two separate distinguishable peaks at different frequencies.    

Finally, we have investigated the impact of the recent re-determination of the $W$ boson mass presented by the CDF collaboration on our conclusions, under the assumption that the inert doublet model be responsible for the observed anomaly. After performing a dedicated scan that accounts for the shift of the Peskin-Takeuchi parameters implied by the new CDF result, in Sec.~\ref{sec:wmass} we have repeated the analysis concerning phase transition patterns and corresponding gravitational wave signals. Whereas all the patterns previously identified remain viable, we find that the number of solution complying with the direct detection bound for dark matter masses above 100 GeV is strongly reduced. For transition chains with at least one first-order step, adopting the CDF determination of $m_W$ forces the bulk of acceptable solutions to fall in the Higgs resonance region, where the $\lambda_{345}$ coupling that regulates also the direct detection cross section is smaller. A similar reduction is observable also for patterns that proceed solely through second-order phase transitions, albeit with reduced intensity. In this case, however, single-step solutions are strongly favoured. We have found no significant difference between the gravitational wave signals implied by this dataset and the results previously obtained by disregarding the anomaly.

\section*{Acknowledgements}

This work was supported by the Estonian Research Council grants PRG434 and PRG356, by the European Regional Development Fund and the programme Mobilitas Pluss grants MOBTT5 and MOBTT86, and by the EU through the European Regional Development Fund CoE program TK133 ``The Dark Side of the Universe''.
The work of LDR has been partially supported by a fellowship from ”la Caixa” Foundation (ID 100010434) and
from the European Union’s Horizon 2020 research and innovation programme under the Marie
Sklodowska-Curie Action grant agreement No 847648.


\bibliographystyle{JHEP}
\bibliography{IDM}

\end{document}